\documentclass{ieeeaccess}

\usepackage{amssymb,amsmath,amsfonts,float,bm}
\usepackage[pdftex]{graphicx}
\usepackage{color}
\usepackage{caption}
\DeclareCaptionFont{ieeeblue}{\color{accessblue}}
\allowdisplaybreaks

\graphicspath{{./Figures/}}
\usepackage{epstopdf}
\usepackage{tabularx}
\usepackage{enumitem}
\usepackage{balance}
\DeclareMathOperator{\tr}{tr}

\DeclareMathOperator{\hcap}{\widehat{\textbf{h}}}
\DeclareMathOperator{\Hcap}{\widehat{\textbf{H}}}
\DeclareMathOperator{\Htilde}{\widetilde{\textbf{H}}}
\newcommand\figwidth{0.44}

\usepackage{cite}
\usepackage{algorithmic}
\usepackage{textcomp}
\begin{document}
\history{Date of publication xxxx 00, 0000, date of current version xxxx 00, 0000.}
\doi{xx.xxxx/ACCESS.2021.DOI}

\title{Beamformer Design with Smooth Constraint-Free Approximation in Downlink Cloud Radio Access Networks}
\author{\uppercase{Fehm\.{I} Emre Kadan}\authorrefmark{1,2}, \IEEEmembership{Student Member, IEEE}, and
\uppercase{Al\.{I} Özgür Yilmaz}\authorrefmark{1}, \IEEEmembership{Member, IEEE}}
\address[1]{Department of Electrical and Electronics Engineering, Middle East Technical University, Ankara 06800, Turkey}
\address[2]{Radar and Electronic Warfare Systems, ASELSAN Inc., Ankara 06830, Turkey}

\markboth
{F. E. Kadan, A. Ö. Yılmaz: Beamformer Design with SCFA in Downlink C-RAN}
{F. E. Kadan, A. Ö. Yılmaz: Beamformer Design with SCFA in Downlink C-RAN}

\corresp{Corresponding author: F. E. Kadan (e-mail: e174115@metu.edu.tr).}

\begin{abstract}
It is known that data rates in standard cellular networks are limited due to inter-cell interference. An effective solution of this problem is to use the multi-cell cooperation idea. In Cloud Radio Access Network, which is a candidate solution in 5G and beyond, cooperation is applied by means of central processors (CPs) connected to simple remote radio heads with finite capacity fronthaul links. In this study, we consider a downlink scenario and aim to minimize total power spent by designing beamformers. We consider the case where perfect channel state information is not available in the CP. The original problem includes discontinuous terms with many constraints. We propose a novel method which transforms the problem into a smooth constraint-free form and a solution is found by the gradient descent approach. As a comparison, we consider the optimal method solving an extensive number of convex sub-problems, a known heuristic search algorithm and some sparse solution techniques. Heuristic search methods find a solution by solving a subset of all possible convex sub-problems. Sparse techniques apply some norm approximation ($\ell_0/\ell_1, \ell_0/\ell_2$) or convex approximation to make the objective function more tractable. We also derive a theoretical performance bound in order to observe how far the proposed method performs off the optimal method when running the optimal method is prohibitive due to computational complexity. Detailed simulations show that the performance of the proposed method is close to the optimal one, and it outperforms other methods analyzed.
\end{abstract}


\titlepgskip=-15pt

\maketitle

\section{Introduction}
In new generation communication systems, the number of devices participating in the network grows exponentially. Furthermore, data rate requirements become challenging to satisfy as the network density increases. Standard cellular systems where a set of user equipments (UEs) are served by a single central base station (BS) have a limited performance due to inter/intra-cell interference. In 5G and beyond, Cloud Radio Access Network (C-RAN) is a candidate solution which uses multi-cell cooperation idea. In C-RAN hierarchy, base stations are simple radio units called remote radio heads (RRHs) which only implement radio functionality such as RF conversions, filtering, and amplifying. All baseband processing is done over a pool of central processors (CPs) which are connected to RRHs with finite capacity fronthaul links. The usage of simple low-power RRHs decreases the cost of deployment as compared to the traditional systems where each BS has its own on-site baseband processor. Furthermore, multi-cell cooperation enables better resource allocation and enhances the performance. In a C-RAN cluster of RRHs and UEs, all transmissions are performed at the same time and frequency band to use the spectrum efficiently.The main architecture of a typical C-RAN system is described in \cite{C-RAN}.

In C-RAN literature, there are two different fronthaul types. The first approach assumes a wired fronthaul where all RRHs are connected to a CP by fiber cables \cite{Wired-SDR}-\hspace{-0.1mm}\cite{Wired-Imperfect-CSI}. In this type of network, there is a natural combinatorial user selection problem. CP should optimize the cooperation of RRHs by determining which user data is sent to which RRHs. In general, the optimal strategy is to try all possible combinations but this approach becomes impractical due to high computational complexity as the network size increases. In the second approach where fronthaul links are wireless, CP transmits user data to RRHs over wireless channels using an antenna array. RRHs operate as a relay to receive and forward the user data to UEs. The relaying mechanism may include decoding of the user data \cite{DF-1}-\hspace{-0.1mm}\cite{DF-2}, or it may be simple amplify-and-forward type relaying \cite{AF-C-RAN}. The first relaying strategy requires user selection to determine which user data to decode as in wired fronthaul networks, however, the second one requires the optimization of the amplifying matrix. 

There are in general three different approaches in the literature to optimize the performance of a downlink C-RAN network. In the first approach, the user data rates are maximized under transmit power and fronthaul capacity constraints \cite{Wired-Rate1}-\hspace{-0.1mm}\cite{Wired-Rate3}. This approach is applicable when each user receives a single data stream or multiple data streams. In the second approach, which is called max-min fairness, the minimum SINR of users is maximized under transmit power constraints \cite{Max-Min-Fairness1}-\hspace{-0.1mm}\cite{Max-Min-Fairness2}. This approach is generally used when each UE has a single antenna and each user receives a single data stream. This type of optimization is applied when the network is power limited. In the last approach, which is called Quality-of-Service (QoS), the total power spent in the system is minimized under user SINR constraints \cite{AF-C-RAN}, \cite{Max-Min-Fairness2}. In this approach, it is guaranteed to satisfy a certain quality of service to each user and the total power spent, which is one of the major costs of an operator, is minimized. 

The techniques used to solve beamforming and user selection problems highly depend on channel estimates and assumptions about the channel estimation errors. There are three different approaches used in the literature about the channel estimation errors. The first approach assumes perfect channel knowledge \cite{Wired-SDR}, \cite{Wired-GreenCRAN}, \cite{DF-1}, \cite{Wired-Rate1}, \cite{Wired-Heuristic1}-\hspace{-0.1mm}\cite{Wired-Heuristic3}. The methods proposed under this assumption may provide some insights but they are not practical especially when the network size is large. In the second approach, the channel estimation errors are assumed to be additive and norm-bounded \cite{AF-C-RAN}, \cite{Max-Min-Fairness2}. This assumption is valid when the quantization error (due to quantization of channel estimates) is dominant. In the last approach, it is assumed that channel estimation errors are additive and their second order statistics are known \cite{Ch-Add}-\hspace{-0.1mm}\cite{ADMM}. This assumption is generally used when major part of error is due to receiver noise in pilot-based channel estimation operation. 

There are a lot of studies existing in the literature related to multi-cell cooperation techniques. The beamforming design and user selection methods involve different techniques. $\ell_0/\ell_1$ norm approximation \cite{l01_1}-\hspace{-0.1mm}\cite{l01_6}, group sparse beamforming with $\ell_0/\ell_2$ norm approximation \cite{Wired-GreenCRAN}, \cite{Wired-Imperfect-CSI}, \cite{GSB}, successive convex approximation \cite{SCA1}-\hspace{-0.1mm}\cite{SCA2}, uplink-downlink duality \cite{UDD}, smooth approximation with subspace projection \cite{Wired-SDR}, \cite{ADMM}, zero-forcing beamforming \cite{ZF}, heuristic search with convex optimization \cite{Wired-Heuristic1}-\hspace{-0.1mm}\cite{Wired-Heuristic3}, difference of convex method \cite{DCF-1}-\hspace{-0.1mm}\cite{DCF-2} are some of the techniques used in the C-RAN system optimization. As another approach, it can be aimed to find the largest set of users which can be served by the RRHs where each user data is sent only by a single RRH \cite{Admission}. The power consumption of RRHs under active and sleeping modes can also be included to the power minimization problem as done in \cite{Wired-GreenCRAN}, \cite{ADMM}. In \cite{Wired-SDR}, the cost function consists of a weighted sum of the total transmit power and the total fronthaul data. Cluster formation \cite{Wired-ClusterFormation}, effect of user traffic delay \cite{Wired-Delay}, and codebook-based designs \cite{SCA2} are also analyzed in the literature. 

In this study, we assume a wired fronthaul network in a downlink C-RAN system. We optimize cooperation and beamforming coefficients for QoS problem under imperfect channel knowledge. We assume that channel estimation is already performed and channel estimates are available. The corresponding channel vectors can be estimated from uplink transmissions in time division duplex systems, however, there is always some estimation error. Hence, in general channel state information (CSI) in CP is imperfect, and the beamforming design algorithms should be robust to channel errors. We assume that channel estimation includes additive errors with known second order statistics. We aim to serve all users in the cluster of interest by minimizing the total power spent in the system. In this study, it is assumed that power has two main components. Firstly, each user data should be first sent by CP to related RRHs. This operation requires some power which can be considered as fronthaul power. Each user data is also sent from RRHs to target users which is the second component of the power in the system. Considering the fact that both fronthaul and RRH power consumptions are significant \cite{ADMM}, we aim to minimize total power spent under per-RRH power transmit constraints by designing beamforming vectors. 

The contributions of the paper can be listed as below: 
\begin{itemize}
\item We define an equivalent combinatorial search problem which can be optimally solved by a finite number of convex optimizations. 
\item The original problem includes discontinuous terms in both cost and constraints. We transform the original problem into a smooth constraint-free function optimization problem and propose a novel method to find a solution using gradient descent. 
\item To make a comparison, we consider a heuristic search method and three sparse techniques with $\ell_0/\ell_1$, $\ell_0/\ell_2$ norm approximations and convex approximation, respectively. 
\item We find a theoretical lower bound for the power spent in terms of channel vectors. The resulting bound can be used for optimality analysis of an algorithm as no algorithm can perform better than the bound. 
\item We perform detailed simulations with realistic channel conditions and compare the performances of all algorithms analyzed. We also make an optimality analysis by comparing the power values of the algorithms with the theoretical lower bound.
\end{itemize}

The organization of the paper is as follows. In Section II, the general system model is described. In Section III, the equivalent problem is presented. Algorithms including smooth approximation, heuristic search approach and sparse techniques are discussed. Some theoretical bounds for the proposed problem are discussed in Section IV. In Section V, simulation results are presented. Finally, Section VI concludes the paper.  

\subsection*{Notation}
Throughout the paper, the vectors are denoted by bold lowercase letters and matrices are denoted by bold uppercase letters. $(\cdot)^T, (\cdot)^H, \tr(\cdot), \text{Re}\{\cdot\}$ indicates the transpose, conjugate transpose, trace and real part operators, respectively. $\textbf{0}$ describes the all-zero matrix, and $\textbf{A} \succeq 0$ implies that the matrix $\textbf{A}$ is Hermitian and positive-semidefinite. $\text{diag}(x_1, x_2, \ldots, x_n)$ denotes the diagonal matrix with diagonal elements $x_1, x_2, \ldots, x_n$ and $\textbf{x}_n$ denotes $n \times n$ diagonal matrix where diagonal entries are all equal to $x$. $\lambda_{\text{max}}(\cdot)$ denotes the maximal eigenvalue of the corresponding square matrix with real eigenvalues. $\mathbb{C}$ denotes the set of complex numbers and $\mathcal{C}\mathcal{N}(\textbf{m}, \textbf{Q})$ denotes the circularly symmetric complex Gaussian random vector with mean $\textbf{m}$ and covariance $\textbf{Q}$. For a vector $\textbf{x}$, $||\textbf{x}||$ denotes the $\ell_2$ norm which is equal to the square root of the sum of absolute squares of the elements and $||\textbf{x}||_0$ denotes the $\ell_0$ norm indicating the number of non-zero elements. Finally, $\delta[\cdot]$ corresponds to the function satisfying $\delta[0]=1, \: \delta[x]=0$ for all $x \neq 0$.

\section{System Model}
We consider a downlink C-RAN scenario where there are $N$ RRHs, which are all connected to a CP via wired fronthaul links, and $K$ users. Each RRH has $L$ antennas and each UE has a single antenna. Our aim is to serve all users with minimum total power spent. All RRH-to-UE channels include both small and large scale fading and they are assumed to be constant over a transmission period. 

\begin{figure}[ht]
\centering
\captionsetup{format=plain, labelsep=period, labelfont={ieeeblue,bf,small}, justification=justified}
\includegraphics[width=0.5\textwidth]{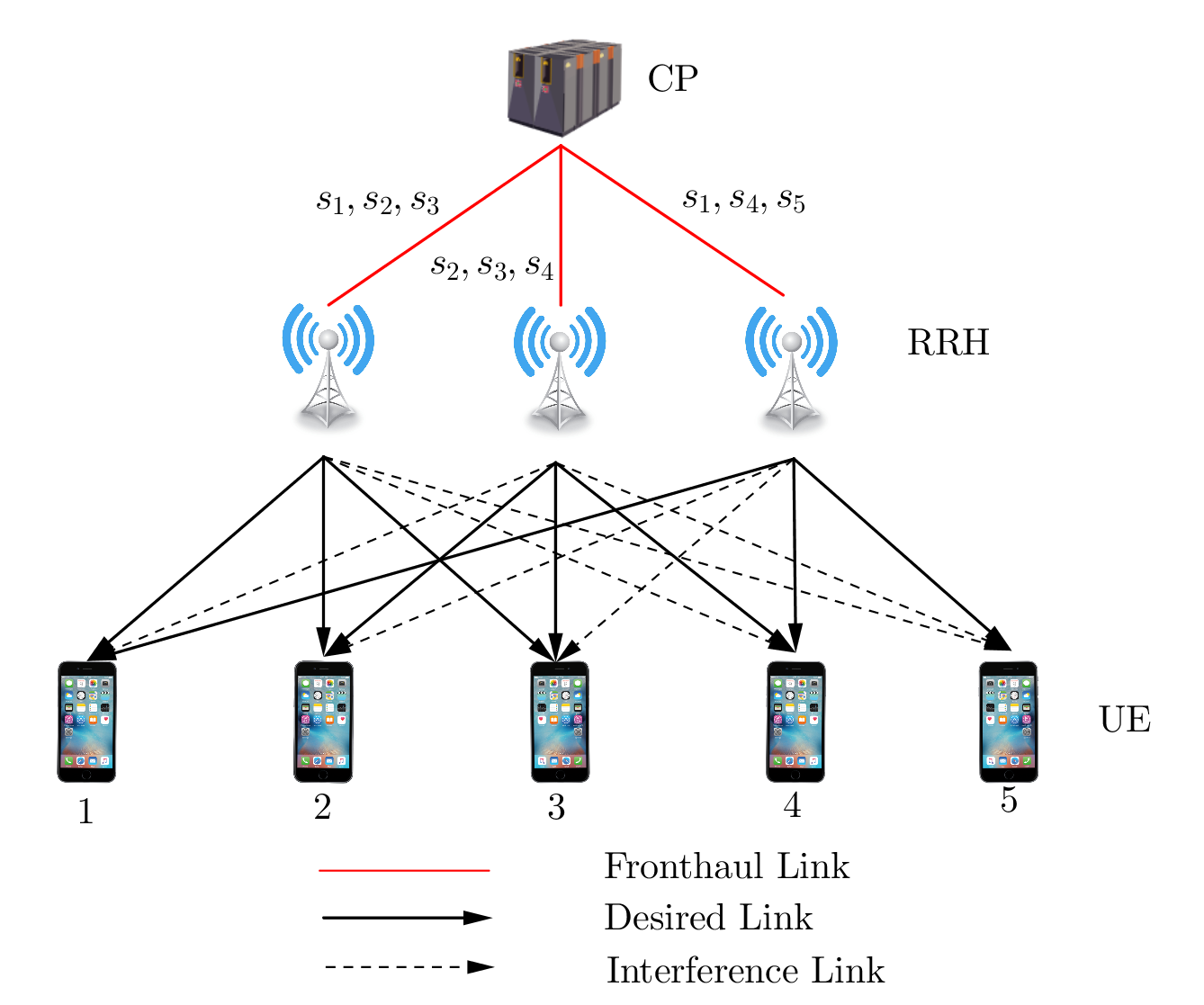}
\caption{An example for block diagram of a downlink C-RAN system with wired fronthaul.}
\end{figure}

In Fig. 1, we see a typical block diagram of a downlink C-RAN with wired fronthaul. In this example, there are $3$ RRHs and $5$ users. Each RRH serves $3$ users. There are $9$ data streams in total transmitted through fronthaul links.

Throughout the paper, the complex baseband model is used. We denote the channel and the corresponding beamformer vectors between the $n$-th RRH and the $k$-th user as $\textbf{h}_{kn} \in \mathbb{C}^L$ and $\textbf{w}_{kn} \in \mathbb{C}^L$, respectively. We assume that there is partial channel knowledge, the channel estimation errors are additive and Gaussian. The second order statistics of the error vectors are known. We use the channel model 
\begin{equation} \label{h_kn}
\textbf{h}_{kn} = \widehat{\textbf{h}}_{kn} + \Delta \textbf{h}_{kn}
\end{equation}
where $\widehat{\textbf{h}}_{kn}$ is the estimated channel vector for the channel between the $n$-th RRH and $k$-th user, and $\Delta \textbf{h}_{kn}$ is the corresponding error with $\Delta \textbf{h}_{kn} \sim \mathcal{C}\mathcal{N}(0, \sigma_{kn}^2 \textbf{I}_L)$. $\Delta \textbf{h}_{kn}$'s are assumed to be independent for all $n, k$. We assume that $\sigma_{kn}$ for all $n, k$ is known by CP. In this study, we assume that CP uses instantaneous channel estimates $\widehat{\textbf{h}}_{kn}$ and error standard deviations $\sigma_{kn}$ to design transmit beamformers. 

If the $n$-th RRH does not send the data of the $k$-th user, then we assume that $\textbf{w}_{kn}=\textbf{0}$. We define augmented channel and beamformer vectors as $\textbf{h}_k=\widehat{\textbf{h}}_{k} + \Delta \textbf{h}_{k}=[\textbf{h}_{k1}^T \hspace{2mm} \textbf{h}_{k2}^T \hspace{2mm} \cdots \hspace{2mm} \textbf{h}_{kN}^T]^T$ and $\textbf{w}_k=[\textbf{w}_{k1}^T \hspace{2mm} \textbf{w}_{k2}^T \hspace{2mm} \cdots \hspace{2mm} \textbf{w}_{kN}^T]^T$. We denote the information data of the $k$-th user by $s_k$ whose power is assumed to be unity. We assume that $n_k \sim \mathcal{C}\mathcal{N}(0, \sigma_k^2)$ where $n_k$ is the noise term in the $k$-th user receiver. In this case, the signal received by the $k$-th user can be expressed by
\begin{align} \label{r_k_1}
r_k &= \displaystyle\sum_{n=1}^{N} \textbf{h}_{nk}^H \displaystyle\sum_{\ell=1}^K \textbf{w}_{n\ell}s_{\ell} + n_k \notag \\
&= \textbf{h}_{k}^H \displaystyle\sum_{\ell=1}^K \textbf{w}_{\ell}s_{\ell} + n_k, \quad \forall k \\
&= \underbrace{\hcap_k^H\textbf{w}_ks_k}_{\text{desired}} +  \underbrace{\Delta \textbf{h}_{k}^H \textbf{w}_ks_k}_{\text{mismatch}} + \underbrace{\textbf{h}_k^H \displaystyle\sum_{\ell \neq k} \textbf{w}_{\ell}s_{\ell}}_{\text{interference}} + \underbrace{n_k}_{\text{noise}}, \: \forall k. \notag \end{align}
In (\ref{r_k_1}), the desired part includes the desired signal for the $k$-th UE. Notice that it contains only the channel estimate for the $k$-th user which is the only useful part for the receiver of corresponding UE. We assume that the effective channel coefficient $\hcap_k^H\textbf{w}_k$ is known by the $k$-th user. The mismatch part is stemming from the channel error for the $k$-th user signal. Although it includes $s_k$ term, the corresponding signal is not useful as its coefficient is not known by the receiver due to uncertainty in the channel estimates. The interference part contains the interference signal including the signals for other users. The noise term is related to UE receiver noise.

Using the equation in (\ref{r_k_1}), we define an SINR for $k$-th user as 
\begin{equation} \label{SINR_1}
\text{SINR}_k = \dfrac{P_d}{P_{m}+P_{\text{int}}+P_{\text{noise}}} 
\end{equation}
where
\begin{equation} \label{P_d}
\begin{aligned}
P_d &= \mathbb{E}[|\hcap_k^H\textbf{w}_ks_k|^2], \: \:
P_{m} = \mathbb{E}[|\Delta \textbf{h}_{k}^H \textbf{w}_ks_k|^2] \\
P_{\text{int}} &= \mathbb{E}[|\textbf{h}_k^H \displaystyle\sum_{\ell \neq k} \textbf{w}_{\ell}s_{\ell}|^2], \: \:
P_{\text{noise}} = \mathbb{E}[|n_k|^2].
\end{aligned}
\end{equation}

Using the fact that $\mathbb{E}\left[s_k^{*}s_{\ell}\right]=\delta[k-\ell]$ and statistics of the channel error and noise terms, we find that\footnote{To calculate the expectations, all channel estimates are assumed to be given and known at the CP which also reveals beamforming vectors to RRHs.} 
\begin{equation} \label{P_d_2}
\begin{aligned}
P_d &= \tr(\Hcap_k\textbf{W}_k), \: P_m = \tr(\textbf{D}_k\textbf{W}_k) \\
P_{\text{int}} &= \displaystyle\sum_{\ell \neq k} \tr(\Htilde_k\textbf{W}_{\ell}), \: P_{\text{noise}} = \sigma_k^2.
\end{aligned}
\end{equation}
where 
\begin{align} \label{H_k} 
\Hcap_k &=\hcap_k\hcap_k^H, \notag \\
\textbf{D}_k &= \mathbb{E}[\Delta\textbf{h}_k\Delta\textbf{h}_k^H] = \text{diag}\left(\sigma_{k1}^2\textbf{I}_L, \sigma_{k2}^2\textbf{I}_L, \ldots, \sigma_{kN}^2\textbf{I}_L\right), \notag \\
\Htilde_k &= \mathbb{E}[\textbf{h}_k\textbf{h}_k^H] = \Hcap_k + \textbf{D}_k, \\
\textbf{W}_k &= \textbf{w}_k\textbf{w}_k^H, \quad \forall k. \notag
\end{align}
Using (\ref{SINR_1}) and (\ref{P_d_2}), we can write
\begin{equation} \label{SINR_2}
\text{SINR}_k = \dfrac{\tr(\Hcap_k\textbf{W}_k)}{\displaystyle\sum_{\ell \neq k} \tr(\Htilde_k\textbf{W}_{\ell}) + \tr(\textbf{D}_k\textbf{W}_k) + \sigma_k^2}, \quad \forall k.
\end{equation}
In Appendix A, we show that the rate $\log_2(1+\text{SINR}_k)$ is achievable for the $k$-th user. Hence, the SINR that we defined can be used as a design criterion. 

A certain SINR is needed to decode the received signals in the UEs. Hence, we compare the SINRs with a pre-determined threshold $\gamma_k$ to decide whether the corresponding user is served. In other words, we assume that $k$-th user is served if and only if $\text{SINR}_k \geq \gamma_k$. If the SINR condition is satisfied, the rate $r_k=\log_2(1+\gamma_k)$ is achievable for $k$-th user. Notice that due to channel estimation errors, the denominator of the SINR expression includes extra positive terms related to $\textbf{D}_k$'s making the SINR smaller compared to the perfect CSI case. Therefore, it is harder to satisfy SINR constraints under imperfect CSI.  

Another design term that can be optimized is the total power spent in the system. The total power $P$ has two components $P_{\text{CP}}$ and $P_{\text{RRH}}$ which correspond to the power consumed by CP and RRHs, respectively. We know that $P_{\text{CP}}$ is an increasing function of the total data rate of users. This is due to the fact that the user data should be first sent by CP to RRHs over fronthaul links before transmitted by RRHs to users. Furthermore, if the cooperation of the RRHs increases, i.e. if the number of RRHs serving a certain user increases, the total fronthaul rate increases resulting in a larger $P_{\text{CP}}$. We use the linear power model as also used in \cite{ADMM}, where $P_{\text{CP}}$ linearly increases with total fronthaul rate. In this model $P_{\text{CP}}$ can be written as 
\begin{equation} \label{P_CP_1}
P_{\text{CP}} = \epsilon_1 r_{\text{tot}} =  \epsilon_1 \displaystyle\sum_{k=1}^K\displaystyle\sum_{n \in S_k} r_k, \\
\end{equation}
where $\epsilon_1$ is a constant multiplier whose unit is $\text{W} \cdot \text{Hz} / \text{bps}$, the total fronthaul rate is denoted by $r_{\text{tot}}$ and $S_k$ is the index set of RRHs serving the user $k$. Using the fact that 
\begin{equation} \label{S_k}
n \in S_k \quad \Longleftrightarrow \quad \textbf{w}_{kn} \neq \textbf{0} \quad \Longleftrightarrow \quad \delta\left[\left\Vert \textbf{w}_{kn} \right\Vert\right] = 0 
\end{equation} 
we get
\begin{equation} \label{P_CP_2}
P_{\text{CP}} = \epsilon_1 \displaystyle\sum_{k=1}^K\displaystyle\sum_{n=1}^N r_k \left(1-\delta\left[\left\Vert \textbf{w}_{kn} \right\Vert\right]\right).
\end{equation} 

The term $P_{\text{RRH}}$ is related to total power transmitted by RRHs which can be written as
\begin{equation} \label{P_tot}
P_{\text{tot}} = \displaystyle\sum_{k=1}^K\displaystyle\sum_{n=1}^N \left\Vert \textbf{w}_{kn} \right\Vert^2.
\end{equation} 
In general, the power spent by RRHs is larger than that of its transmit power due to inefficiency of power amplifiers. We assume that power spent by RRHs is equal to $P_{\text{RRH}}  = \epsilon_2  P_{\text{tot}}$ where $\epsilon_2$ is a unitless constant multiplier.

We know that there is a natural power transmit constraint for each RRH, which can be formulated as 
\begin{equation} \label{P_t}
P_n = \displaystyle\sum_{k=1}^K \Vert \textbf{w}_{kn} \Vert^2 \leq P_t, \quad \forall n
\end{equation} 
where $P_n$ is the transmit power of $n$-th RRH and $P_t$ is the power transmit threshold value which is assumed to be the same for all RRHs. 

In this study, we aim to serve all $K$ users by minimizing $P=P_{\text{CP}} + P_{\text{RRH}}$ under per-RRH power transmit constraints. Considering all constraints, the design problem can be given by 
\begin{align} \label{P1}
\text{(P1)} \quad & \min_{\textbf{w}_{kn}} \: \: \displaystyle\sum_{k=1}^K\displaystyle\sum_{n=1}^N \epsilon_1 r_k \left(1-\delta\left[\left\Vert \textbf{w}_{kn} \right\Vert\right]\right) + \epsilon_2 \left\Vert \textbf{w}_{kn} \right\Vert^2 \notag \\
\text{s.t.} \: \: & \displaystyle\sum_{k=1}^K \Vert \textbf{w}_{kn} \Vert^2  \leq P_t, \quad \forall n=1, 2, \ldots, N, \\
&\text{SINR}_k \geq \gamma_k, \quad \forall k=1, 2, \ldots, K. \notag
\end{align}

\section{Algorithms}
In this section, we present the algorithms that we use while solving (P1). Firstly, we define an equivalent combinatorial search problem. We propose a novel method that transforms the problem into a smooth constraint-free form and solve it using gradient descent. The initial point of this method is found from the equivalent problem. As a comparison we also consider a heuristic search method that solves the equivalent problem using successive convex optimizations, and three sparse algorithms using $\ell_0/\ell_1$ norm approximation, $\ell_0/\ell_2$ norm approximation and convex approximation, respectively. 

\subsection{Equivalent Combinatorial Search Problem}
(P1) can be transformed into a combinatorial search problem which can be optimally solved with a finite number of convex optimizations. Firstly, to analyze the cooperation between the RRHs, we define a network link matrix $\textbf{C}$ with dimensions $K \times N$, where each entry $c_{kn}$ is either $0$ or $1$. If the $n$-th RRH sends the $k$-th user data, then we say that there is a link between the $n$-th RRH and the $k$-th user and set $c_{kn}=0$. Otherwise $c_{kn}=1$. Similarly, we can relate the entries of $\textbf{C}$ with the index sets $S_k$ as 
\begin{equation} \label{C_vs_S_k}
n \in S_k \quad \Longleftrightarrow \quad c_{kn}=0.
\end{equation}
Using this definition, we can write a relation between beamformers and entries of $\textbf{C}$ as
\begin{equation} \label{c_kn}
\displaystyle\sum_{n=1}^N c_{kn} \textbf{B}_n \textbf{w}_k = \textbf{0}, \quad \forall k \end{equation}
where $\textbf{B}_n$ is an $NL \times NL$ diagonal matrix defined as 
\begin{equation} \label{B_n}
\textbf{B}_n = \text{diag}\left(\textbf{0}_{(n-1)L}, \: \textbf{1}_L, \: \textbf{0}_{(N-n)L}\right), \: n=1,2, \ldots, N. \end{equation} 
The matrices $\displaystyle\sum_{n=1}^N c_{kn} \textbf{B}_n$ are diagonal with diagonal entries $0$ and $1$, and hence they are all positive semi-definite. Therefore, it follows that
\begin{equation} \label{Cond1} 
\displaystyle\sum_{n=1}^N c_{kn} \textbf{B}_n \textbf{w}_k = \textbf{0} \: \Longleftrightarrow \: \displaystyle\sum_{n=1}^N c_{kn} \tr\left(\textbf{B}_n \textbf{W}_k\right) = 0, \quad \forall k. \end{equation}
Using the $\textbf{B}_n$ matrices defined, the total transmit power of the $n$-th RRH can be written as 
\begin{equation} \label{P_n} 
P_n = \displaystyle\sum_{k=1}^K \Vert \textbf{w}_{kn} \Vert^2 = \displaystyle\sum_{k=1}^K \tr(\textbf{B}_n \textbf{W}_k). \end{equation} 
(P1) is a problem involving terms related to the number of zero beamformers which are discontinuous functions of $\textbf{w}_{kn}$'s. To find a solution, we can consider the problem for a fixed $\textbf{C}$ which is denoted by (P2). If we solve (P2) for all possible $\textbf{C}$ matrices, then we can find the optimal solution of (P1) by finding the optimal $\textbf{C}$ that minimizes $P$. When we fix $\textbf{C}$, the term $P_{\text{CP}}$ becomes also fixed, and the problem reduces to finding optimal beamformers minimizing the total transmit power from RRHs-to-UEs. Using (\ref{c_kn})-(\ref{P_n}), (P2) can be expressed by
\begin{align} 
\label{Cost} \text{(P2)} & \: \min_{\{\textbf{W}_k\}_{k=1}^K} \hspace{2mm} \displaystyle\sum_{k=1}^K \tr(\textbf{W}_k) \\
\label{Constraint1} \text{s.t.} \quad & \dfrac{\tr(\Hcap_k\textbf{W}_k)}{\displaystyle\sum_{\ell \neq k} \tr(\Htilde_k\textbf{W}_{\ell}) + \tr(\textbf{D}_k\textbf{W}_k) + \sigma_k^2} \geq \gamma_{k} \: \: \forall k, \\  
\label{Constraint2} & \displaystyle\sum_{n=1}^N c_{kn} \tr\left(\textbf{B}_n \textbf{W}_k\right) = 0 \quad \forall k, \\ 
\label{Constraint3} & \displaystyle\sum_{k=1}^K \tr(\textbf{B}_n \textbf{W}_k) \leq P_t \quad \forall n, \\ 
\label{Constraint4} & \text{rank} (\textbf{W}_k) = 1, \textbf{W}_k \succeq 0 \quad \forall k. \end{align}
In (\ref{Constraint1}) the SINR constraints are given. (\ref{Constraint2}) is the relation between $\textbf{C}$ matrix and beamformers. (\ref{Constraint3}) is the per-RRH power constraint. Finally, (\ref{Constraint4}) includes a rank constraint which makes the problem non-convex. Using a similar method as discussed in \cite{Wired-SDR}, it can be shown that (P2) is NP-hard. As done in \cite{Wired-SDR}, \cite{Wired-GreenCRAN}, we can make a semi-definite relaxation by omitting the rank constraint. In the relaxed problem, cost and constraints are convex and hence standard convex optimization methods can be applied to find a solution. As a result, in order to find a solution of (P1), we need to solve (P2) for all possible $\textbf{C}$ matrices and determine the one with the minimum $P$ value. This is a combinatorial search problem that requires a lot of convex optimizations. In order to find a practical solution, one way is to apply a heuristic search over $\textbf{C}$ matrices. In general, the performance of this approach enhances as the number of trials increases. Another way is to apply a direct approximation on the variables in cost and constraints of the problem which is discussed in subsection C. 

\subsection{Rank-1 Approximation}
In (P2), we find a solution by omitting the rank constraint. If the result is rank-1, the solution becomes optimal. In Appendix B, using the idea in \cite{SCA1}, we prove that the optimal solution of the relaxed problem is always rank-1. Hence by applying eigenvalue decomposition and taking the principal eigenvalue and eigenvector, we find the optimal solution of (P2). If (P2) is solved for all possible $\textbf{C}$ matrices, then we can find the optimal solution of (P1).

\subsection{Smooth Constraint-Free Approximation (SCFA)}
The combinatorial method involving a series of convex optimizations for relaxed version of (P2) can be used to find the optimal solution of (P1), however, the computational complexity is very high when the numbers $K, N, L$ are large. To find a practical solution, we propose a new method which is based on approximating discontinuous functions with a sequence of smooth function sequences. \cite{Wired-SDR}, \cite{ADMM} also use a similar idea, but there is an additional constraint which requires a subspace projection at each iteration step. In the related papers, authors only approximate the cost function to be optimized while the constraints are kept the same. Their solution requires an extra subspace projection step. They first optimize an approximated cost function where constraints are not considered. After this step, the solution is projected onto the space defined by the constraints. In our method, we shift all constraints to the main cost function to be optimized using a similar approximation idea. Hence, standard gradient search methods can be applied to the final constraint-free smooth function of beamformer vectors. By this way, we can directly optimize the approximated cost function without any extra subspace projection operations.

Firstly, we prove a theorem about the SINR terms for the optimal solution: 

\noindent \textbf{Theorem 1:} For the optimal solution of (P1), we have
\begin{center} $\text{SINR}_k = \gamma_k, \: \: \forall k. $ \end{center}
\textbf{Proof:} The proof is given in Appendix C. 

\noindent Using Theorem 1, we prefer to express SINR constraints as $\text{SINR}_k = \gamma_k$, i.e., $\rho_k=0, \: \forall k$ where $\rho_k$ is defined as
\begin{equation} \label{rho}
\rho_k =  \tr\left((\Hcap_k-\gamma_k\textbf{D}_k)\textbf{W}_k\right) - \gamma_k\displaystyle\sum_{\ell \neq k} \tr(\Htilde_k\textbf{W}_{\ell}) - \gamma_k\sigma_k^2. \end{equation}
As we see in (\ref{P_CP_2}), the term related to $P_{\text{CP}}$ is discontinuous with respect to $\textbf{w}_{kn}$'s, but can be expressed as the limit of some smooth function sequences. For this purpose, we use the fact
\begin{equation} \label{Fact 1}
\text{Fact 1:} \quad  e^{-\frac{x^2}{a}} \to \delta[x] \quad \text{as} \quad a \to 0^{+}. \end{equation}
Using Fact 1, we can approximate $P_{\text{CP}}$ as 
\begin{equation} \label{P_CP_3}
P_{\text{CP}} \approx \epsilon_1 \displaystyle\sum_{k=1}^K\displaystyle\sum_{n=1}^N r_k\left(1-\exp\left(-\frac{\left\Vert \textbf{w}_{kn} \right\Vert^2}{a}\right)\right) 
\end{equation}
where $a$ is a small positive number. Using (\ref{P_CP_3}), the approximate expression for $P$ becomes smooth with respect to beamforming vectors. We use a similar way to shift the constraints to the main expression. Notice that
\begin{equation} \label{Fact 2}
\text{Fact 2:} \quad  e^{\frac{x}{a}} \to  
\begin{cases}
    0, & \text{if} \quad x < 0 \\
    \infty, & \text{if} \quad x > 0
\end{cases} \quad \text{as} \quad a \to 0^{+}. \end{equation}
\begin{equation} \label{Fact 3} 
\text{Fact 3:} \quad  e^{\frac{x^2}{a}}-1 \to  
\begin{cases}
    0, & \text{if} \quad x = 0 \\
    \infty, & \text{if} \quad x \neq 0
\end{cases} \quad \text{as} \quad a \to 0^{+}. \end{equation}
Consider the expressions 
\begin{equation} \label{c_12}
\begin{aligned}
c_1 &= \displaystyle\sum_{n=1}^N \exp\left(\frac{1}{a}\left(\displaystyle\sum_{k=1}^K \left\Vert \textbf{w}_{kn} \right\Vert^2 - P_t \right)\right), \\
c_2 &= \displaystyle\sum_{k=1}^K \left[\exp\left(\dfrac{\rho_k^2}{a}\right)-1\right].
\end{aligned}
\end{equation}
Using Fact 2, $c_1$ tends to $0$ if the per-RRH power transmit constraints are satisfied, and tends to infinity otherwise. Similarly, by Fact 3, $c_3$ tends to $0$ if SINR constraints are satisfied, and tends to infinity otherwise. Therefore, we can delete the constraints and express the problem as 
\begin{align}
P &= \epsilon_1 r_{\text{tot}} + \epsilon_2 P_{\text{tot}} \approx  Q \notag \\
&= \displaystyle\sum_{k=1}^K\displaystyle\sum_{n=1}^N \epsilon_1 r_k\left(1-\exp\left(-\frac{\left\Vert \textbf{w}_{kn} \right\Vert^2}{a}\right)\right)+\epsilon_2 \left\Vert \textbf{w}_{kn} \right\Vert^2 \notag \\
\label{Q} &+ \eta_1\displaystyle\sum_{n=1}^N \exp\left(\frac{1}{a}\left(\displaystyle\sum_{k=1}^K \left\Vert \textbf{w}_{kn} \right\Vert^2 - P_t \right)\right) \\
&+ \eta_2 \displaystyle\sum_{k=1}^K\left[\exp\left(\frac{\rho_k^2}{a}\right)-1\right]\notag
\end{align}
where $a>0$ is a small number\footnote{In fact, we can use different $a$'s for approximating $P_{\text{CP}}$ and for constraint shifting. However, to make the expression simpler, we use the same $a$ for all three approximations.}, and $\eta_1, \eta_2$ are two positive weights. Notice that using (\ref{Q}), we shift the constraints to the main expression. When the constraints in (\ref{P1}) is satisfied, corresponding terms in (\ref{Q}) becomes $0$ and does not affect the value of $Q$. If they are not satisfied, we get $Q \to \infty$. So, once $Q$ is minimized, the constraints should be automatically satisfied. Note that using this approach, $Q$ becomes a smooth function in terms of $\textbf{w}_{kn}$'s and hence standard gradient search algorithms can be applied. We can directly minimize $Q$ without any constraint using the SCFA algorithm. 

In SCFA, we start from an initial point and update the beamformers iteratively by evaluating the gradient. In the update equation, we use the augmented beamformer $\textbf{w}=[\textbf{w}_1^T \textbf{w}_2^T \cdots \textbf{w}_K^T]^T$. The equation in (\ref{Q}) can be written in terms of $\textbf{w}$ as given in (\ref{Q_2}). 
\begin{align} 
Q &= \epsilon_1 \displaystyle\sum_{k=1}^K\displaystyle\sum_{n=1}^N r_k\left[1-\exp\left(-\frac{\textbf{w}^H \textbf{B}_{kn} \textbf{w}}{a}\right)\right]+\epsilon_2 \textbf{w}^H \textbf{w} \notag \\
\label{Q_2} &+ \eta_1\displaystyle\sum_{n=1}^N \exp\left(\frac{\textbf{w}^H \textbf{C}_n \textbf{w} - P_t}{a}\right) \\ 
&+ \eta_2 \displaystyle\sum_{k=1}^K \left[\exp\left(\frac{\left(\textbf{w}^H \textbf{A}_k \textbf{w} - d_k\right)^2}{a}\right)-1\right] \notag 
\end{align}
where 
\begin{align} 
\textbf{A}_k &= \text{diag}\big(-\gamma_k \widetilde{\textbf{H}}_k, \: \ldots, \: -\gamma_k \widetilde{\textbf{H}}_k, \: \overbrace{\Hcap_k-\gamma_k\textbf{D}_k}^{k\text{-th}}, \: -\gamma_k \widetilde{\textbf{H}}_k, \notag \\
&\ldots, -\gamma_k \widetilde{\textbf{H}}_k\big) \notag \\
\label{A_k} \textbf{B}_{kn} &= \text{diag}\left(\textbf{0}_{\left((k-1)N+n-1\right)L}, \: \textbf{1}_{L}, \: \textbf{0}_{\left((K-k+1)N - n\right)L}\right) \\
\textbf{C}_{n} &= \displaystyle\sum_{k=1}^K \textbf{B}_{kn} = \text{diag}(\overbrace{\textbf{B}_n, \textbf{B}_n, \ldots, \textbf{B}_n}^{K}), \quad d_k = \sigma_k^2\gamma_k \notag
\end{align}
for all $n=1, 2, \ldots, N$ and $k=1, 2, \ldots, K$. In this case, the update equation can be given as 
\begin{equation} \label{gradient_descent} \textbf{w}^{(t+1)} =  \textbf{w}^{(t)} - \mu^{(t)} \bm{\nabla}(\textbf{w}^{(t)}). \end{equation} 
Here $t$ denotes the iteration index, $\mu^{(t)}$ is the step-size, $\bm{\nabla}(\textbf{w}^{(t)})$ is the gradient vector which can be calculated through (\ref{Q_2}).

We know that a small step-size causes slow convergence and large step-size implies divergence. To obtain a suitable step-size, we use a well-known variable step-size Barzilai-Borwein (BB) method \cite{BB}, which calculates the step-size as 
\begin{equation} \label{step-size1} \mu^{(t)} = \frac{\left(\delta\bm{\nabla}^{(t)}\right)^H \delta\textbf{w}^{(t)}}{\left(\delta\bm{\nabla}^{(t)}\right)^H \delta\bm{\nabla}^{(t)}}  \end{equation} 
where 
\begin{equation} \label{step-size2}
\begin{aligned}
\delta\bm{\nabla}^{(t)} &= \bm{\nabla}(\textbf{w}^{(t)}) - \bm{\nabla}(\textbf{w}^{(t-1)}), \\
\quad \delta\textbf{w}^{(t)} &= \textbf{w}^{(t)} - \textbf{w}^{(t-1)}. 
\end{aligned}
\end{equation} 
In SCFA, the choice of initial point $\textbf{w}^{(0)}$ is crucial to obtain a good convergence. We know that for a given link configuration, the problem can be optimally solved using convex optimization. Using this idea, we choose the initial point as the solution of the full cooperation case. We operate the algorithm until the rate of change of $Q$ is small enough. In order to obtain a faster convergence, we change the value of $a$ throughout the iterations. If the rate of change of $Q$ is small enough, we decrease $a$ by some factor. 

After convergence, we obtain a solution which satisfies a local minimum for $Q$. On the other hand, because of the approximation done, the value of $Q$ may not be exactly equal to $P$. Therefore, we first find the link configuration and the corresponding $\textbf{C}$ matrix according to the $\ell_2$ norms of the beamformers obtained at the last step of the iteration, then solve the problem using convex optimization for the matrix $\textbf{C}$ found. After the convergence, some beamformers become very close to zero which means that the corresponding links should not be used. The steps are listed in Algorithm below. 

\noindent\rule{0.5\textwidth}{0.8pt} 
\vspace{2mm}
\textbf{Algorithm} (Smooth Constraint-Free Approximation, SCFA) \vspace{-4.5mm} \\
\noindent\rule{0.5\textwidth}{0.4pt} \\
\textbf{Step 1, Initialization:} 
\begin{itemize}[leftmargin=0.2in, label={}]
\item Solve (P2) for full cooperation case. Set the initial value $\textbf{w}^{(0)}$. Define $\mu^{(1)} = 10^{-4}, t_{\text{max}} = 10^5, \tau=10^{-6}, \xi = 0.1, Q_t = 10^{-6}, \kappa=10^{-2} , \Delta = 5, \eta_1 = \eta_2 = (\epsilon_1+\epsilon_2)/2$. 
\end{itemize}
\textbf{Step 2, Gradient Descent:} 
\begin{itemize}[leftmargin=0.2in, label={}]
\item For $t=1, 2, \ldots, t_{\text{max}}$ repeat the following steps:
\item 1) Find the gradient $\bm{\nabla}(\textbf{w}^{(t)})$ and step-size $\mu^{(t)}$. 
\item 2) Update the beamformers: 
\begin{equation*} \textbf{w}^{(t+1)} = \textbf{w}^{(t)} - \mu^{(t)} \bm{\nabla}(\textbf{w}^{(t)}). \end{equation*} 
\item 3) Check the condition for $a$: 
\begin{equation*} a^{(t+1)} =  
\begin{cases}
    \xi a^{(t)}, & \text{if} \: \: |Q^{(t+1)}-Q^{(t)}|<\tau a^{(t)} \\
    a^{(t)}, & \text{otherwise}
\end{cases} \end{equation*}
\item 4) Check the condition for termination:
\begin{center} If $\underset{t-\Delta \leq t_s \leq t}{\max}|Q^{(t+1)}-Q^{(t_s)}|<Q_t$, then terminate. \end{center}
\item After the termination, evaluate the $\ell_2$ norms of the final beamformer vectors. Make a priority list of pairs $(k, n)$ so that $(k_1, n_1)$ is more prior than $(k_2, n_2)$ if and only if $\|\textbf{w}_{k_1n_1}\|>\|\textbf{w}_{k_2n_2}\|$. 
\end{itemize}
\textbf{Step 3, Find the link configuration:} 
\begin{itemize}[leftmargin=0.2in, label={}]
\item Form a network link matrix $\textbf{C}$ such that
\begin{equation*} c_{kn} =  
\begin{cases}
    1, &\hspace{-2mm}\text{if} \: \: \left\Vert \textbf{w}_{kn} \right\Vert < \kappa P_t \\
    0, &\hspace{-2mm}\text{otherwise}
\end{cases} \end{equation*}
\item Repeat the following steps until the problem becomes feasible: 
\item 1) Solve (P2) using $\textbf{C}$ with convex optimization. If the problem is feasible, then stop. 
\item 2) If it is not feasible, find the most prior pair $(k, n)$ with $c_{kn}=1$. Update $\textbf{C}$ by making $c_{kn}=0$. \footnote{There may be some cases where direct removal of links with small norm results in an infeasible problem. Hence, we iteratively add a new link until the problem becomes feasible.} 
\end{itemize}
\vspace{-3mm}
\noindent\rule{0.5\textwidth}{0.4pt} 

In Step 1, we find the initial point to start the algorithm. In Step 2, iterative gradient descent process is operated. In the final step, the network link matrix is determined and the final solution is found by a small number of convex optimizations. 

\subsection{Heuristic Search Algorithms}
We know that the optimal solution of (P1) can be found by solving (P2) for all possible $\textbf{C}$ matrices. In Theorem 2, we give the total number of such $\textbf{C}$ matrices. 

\noindent \textbf{Theorem 2:} The number of all possible $\textbf{C}$ matrices is given by 
\begin{equation} \label{N_C} N_{\textbf{C}} = (2^N-1)^K. \end{equation}
\textbf{Proof:} The proof is given in Appendix D. 

\noindent As we see in Theorem 2, the number of convex optimizations required grows exponentially with $N$ and $K$. Therefore, we can apply a heuristic search over $\textbf{C}$ matrices to find a practical solution. In this study, we consider a heuristic search method based on successive link removal. To make a comparison, we can also apply the exhaustive search method which tries all possible $\textbf{C}$ matrices. In all these methods, we try some subset of all possible $\textbf{C}$ matrices and find the solution among them corresponding to the minimum $P$ value. \\

\noindent \textbf{Method 1: Iterative Link Removal (ILR)} \\
This algorithm is discussed in \cite{Wired-Heuristic1} for a problem where the total power is limited and the main concern is to minimize the cooperation among the RRHs. According to the beamformers found in the full cooperation case, corresponding entries of $\textbf{C}$ are prioritized. The entries with less priority are iteratively equalized to $1$ until the problem becomes infeasible. Since there are $NK$ links in total and each user should have at least one link, this algorithm requires at most $NK-K$ convex optimizations. \\

\noindent \textbf{Method 2: Exhaustive Search (ES)} \\
This method optimally solves (P1) by trying all $(2^N-1)^K$ possible $\textbf{C}$ matrices. As there are a lot of convex optimizations required, this method is not practical and we use it as a comparison only for some small values of $K$ and $N$.

\subsection{Sparse Algorithms} 
The objective function of (P1) includes terms related to the number of zero beamformers. The corresponding terms can be written as $\ell_0$ norms of some vectors. $\ell_0$ norm can be approximated by $\ell_1$ norm \cite{l01_1}-\hspace{-0.1mm}\cite{l01_6}, $\ell_2$ norm \cite{Wired-GreenCRAN}, \cite{Wired-Imperfect-CSI}, \cite{GSB} or some convex function \cite{SCA1}-\hspace{-0.1mm}\cite{SCA2}. \\        

\noindent \textbf{Method 1: Majorization-Minimization (MM)} \\
This method uses the idea of $\ell_0/\ell_1$ norm approximation. In \cite{l01_1}-\hspace{-0.1mm}\cite{l01_6}, this approximation is applied for similar problems.  For a non-negative real number $x$, we have
\begin{equation} \label{l01_eq1}
\lim_{\theta \to 0} \dfrac{\ln(1+\theta^{-1}x)}{\ln(1+\theta^{-1})} = 
\begin{cases}
    1, &\hspace{-2mm}\text{if} \: \: x>0 \\
    0, &\hspace{-2mm}\text{if} \: \: x=0.
\end{cases} \end{equation}
Using the observation given in (\ref{l01_eq1}), we can approximate $\ell_0$ norm by 
\begin{equation}
||x||_0 \approx c_{\theta} \cdot \ln(1+\theta^{-1}x)
\end{equation}
where $\theta>0$ is a small number and $c_{\theta}=\dfrac{1}{\ln(1+\theta^{-1})}$ is a constant. The function $\ln(1+\theta^{-1}x)$ is concave with respect to $x$ and hence it is upper bounded by its first order Taylor series expansion, i. e.,
\begin{equation}
\ln(1+\theta^{-1}x) \leq \dfrac{x}{x+\theta}.
\end{equation}
Therefore, $||x||_0$ can be approximated by $c_{\theta} \cdot \dfrac{x}{x+\theta}$ for any non-negative real number $x$. For a vector $[x_1 \: x_2 \: \cdots \: x_n]$ with non-negative real elements, we obtain that
\begin{equation}
\lVert [x_1 \: x_2 \: \cdots \: x_n]\lVert_0 \approx c_{\theta} \cdot \displaystyle\sum_{m=1}^n \dfrac{x_m}{x_m+\theta}. 
\end{equation}
This method is called majorization-minimization (MM) method as the upper bound of the $\ell_0$ norm will be minimized. To use this method in our problem, we first write the discontinuous term in the objective function in terms $\ell_0$ norms. 
\begin{equation}
\displaystyle\sum_{n=1}^N\left(1-\delta[||\textbf{w}_{kn}||]\right) = \lVert [\lVert\textbf{w}_{k1}\lVert^2 \: \lVert\textbf{w}_{k2}\lVert^2 \:  \cdots \: \lVert\textbf{w}_{kN}\lVert^2] \lVert_0.
\end{equation}
We approximate the $\ell_0$ norm as
\begin{equation}
\begin{aligned}
\displaystyle\sum_{n=1}^N\left(1-\delta[||\textbf{w}_{kn}||]\right) &\approx c_{\theta}\displaystyle\sum_{n=1}^N \dfrac{\textbf{w}_{kn}^H\textbf{w}_{kn}}{\textbf{w}_{kn}^H\textbf{w}_{kn}+\theta} \\
&= c_{\theta}\displaystyle\sum_{n=1}^N \dfrac{\tr(\textbf{B}_n\textbf{W}_k)}{\tr(\textbf{B}_n\textbf{W}_k)+\theta}.
\end{aligned}
\end{equation}
In this case, the total power can be written as
\begin{equation}
P \approx \epsilon_1 c_{\theta} \displaystyle\sum_{k=1}^K r_k \displaystyle\sum_{n=1}^N \dfrac{\tr(\textbf{B}_n\textbf{W}_k)}{\tr(\textbf{B}_n\textbf{W}_k)+\theta} + \epsilon_2 \displaystyle\sum_{k=1}^K \tr(\textbf{W}_k).
\end{equation}
To obtain a convex objective function, as done in \cite{l01_1}-\hspace{-1mm}\cite{l01_6}, we optimize $P$ iteratively. We approximate the term  $\dfrac{\tr(\textbf{B}_n\textbf{W}_k)}{\tr(\textbf{B}_n\textbf{W}_k)+\theta}$ as $\dfrac{\tr(\textbf{B}_n\textbf{W}_k)}{\tr(\textbf{B}_n\textbf{W}_k^{(t)})+\theta}$ where $\textbf{W}_k^{(t)}$ is the solution found at the previous iteration. After the final approach, the expression of $P$ becomes linear (and hence convex) with respect to $\textbf{W}_k$'s and the optimization can be performed by omitting the rank-1 constraint of $\textbf{W}_k$'s. In the first part of Appendix E, we show that the optimal solution of the relaxed problem is always rank-1. The corresponding problem can be expressed as
\begin{align} \label{P3}
\text{(P3)} \: &\min_{\textbf{W}_{k}} \: \displaystyle\sum_{k=1}^K \left(r_k \displaystyle\sum_{n=1}^N \dfrac{\epsilon_1 c_{\theta} \tr(\textbf{B}_n\textbf{W}_k)}{\tr(\textbf{B}_n\textbf{W}_k^{(t)})+\theta} + \epsilon_2\tr(\textbf{W}_k)\right), \notag \\
\text{s.t.} \: \: & \displaystyle\sum_{k=1}^K \tr(\textbf{B}_n\textbf{W}_{k}) \leq P_t, \quad \forall n=1, 2, \ldots, N, \\
&\text{SINR}_k \geq \gamma_k, \quad \textbf{W}_k \succeq 0 \quad \forall k=1, 2, \ldots, K. \notag
\end{align}
In (\ref{P3}), the terms $\textbf{W}_k^{(t)}$'s are the solutions found at the previous iteration. The steps of MM algorithm are listed below.

\noindent\rule{0.5\textwidth}{0.8pt} 
\vspace{2mm}
\textbf{Algorithm} (Majorization-Minimization, MM) \vspace{-4.5mm} \\
\noindent\rule{0.5\textwidth}{0.4pt} \\
\textbf{Step 1, Initialization:} 
\begin{itemize}[leftmargin=0.2in, label={}]
\item Solve (P2) for full cooperation case. Set the initial values of $\textbf{W}_k^{(0)}$ for $k=1, 2, \ldots, K$. Define $\theta=10^{-5}, \: t_{\text{max, 1}}=10^3, \: \Delta_1=10^{-5}, \: \kappa_1 = 10^{-3}$.
\end{itemize}
\textbf{Step 2, Successive Convex Optimizations:} 
\begin{itemize}[leftmargin=0.2in, label={}]
\item For $t=1, 2, \ldots, t_{\text{max, 1}}$ repeat the following steps:
\item 1) Solve (P3) using $\textbf{W}_k^{(t-1)}$'s found in the previous iteration. 
\item 2) Check the condition for termination:
\begin{center} If $|P^{(t)}-P^{(t-1)}|/P^{(t)}<\Delta_1$, then terminate. \end{center}
\item After the termination, evaluate the $\ell_2$ norms of the final beamformer vectors. 
\end{itemize}
\textbf{Step 3, Find the link configuration:} 
\begin{itemize}[leftmargin=0.2in, label={}]
\item 1) Form a network link matrix $\textbf{C}$ by removing the links with small norms, i.e., $\left\Vert \textbf{w}_{kn} \right\Vert < \kappa_1 P_t$.
\item 2) Solve (P2) using $\textbf{C}$ with convex optimization. 
\end{itemize}
\vspace{-3mm}
\noindent\rule{0.5\textwidth}{0.4pt} \\

\noindent \textbf{Method 2: Group Sparse Beamforming (GSB)} \\
In this algorithm, we use $\ell_0/\ell_2$ norm approximation which corresponds to a homogeneous convex lower bound \cite{Wired-GreenCRAN}, \cite{Wired-Imperfect-CSI}, \cite{GSB}. Firstly, we define 
\begin{equation}
f(k, n)=\epsilon_1r_k(1-\delta[\lVert \textbf{w}_{kn} \lVert])+\epsilon_2\lVert\textbf{w}_{kn}\lVert^2, \quad \forall k, n.
\end{equation}
We can write the total power spent $P$ as 
\begin{equation}
P=\displaystyle\sum_{k=1}^K\displaystyle\sum_{n=1}^N f(k, n). 
\end{equation} 
Using the fact that
\begin{equation}
f(k, n)= 
\begin{cases}
    0, &\hspace{-2mm}\text{if} \: \: \textbf{w}_{kn} = \textbf{0} \\
    \epsilon_1r_k+\epsilon_2\lVert\textbf{w}_{kn}\lVert^2, &\hspace{-2mm}\text{if} \: \: \textbf{w}_{kn} \neq \textbf{0},
\end{cases} \end{equation}
and by the well-known Arithmetic-Geometric Mean Inequality, we obtain that
\begin{equation}
f(k, n) \geq 2\sqrt{\epsilon_1\epsilon_2r_k}\lVert\textbf{w}_{kn}\lVert, \quad \forall k, n.
\end{equation} 
It follows that
\begin{equation} \label{GSB_eq1}
P \geq \displaystyle\sum_{k=1}^K\displaystyle\sum_{n=1}^N 2\sqrt{\epsilon_1\epsilon_2r_k}\lVert\textbf{B}_n\textbf{w}_{k}\lVert.
\end{equation} 
The right-hand side of (\ref{GSB_eq1}) is homogeneous and convex with respect to beamformer vectors. In this method, we optimize the convex lower bound for total power spent given in (\ref{GSB_eq1}) under the SINR and per-RRH transmit power constraints. We can write both two constraints as second order cone constraints. Firstly, we can express SINR constraints as
\begin{equation} \label{GSB_eq2}
(1+\gamma_k)\textbf{w}_k^H\Hcap_k\textbf{w}_k \geq \gamma_k\displaystyle\sum_{\ell=1}^{K} \textbf{w}_{\ell}^H\Htilde_k\textbf{w}_{\ell} + \gamma_k\sigma_k^2, \quad \forall k.
\end{equation} 
As the matrices $\Htilde_k$'s are positive semi-definite for all $k$, using the Cholesky factorization, we can find matrices $\textbf{T}_k$ such that $\Htilde_k=\textbf{T}_k^H\textbf{T}_k$ for all $k$. Using this fact, we can rewrite (\ref{GSB_eq2}) as 
\begin{equation}
(1+\gamma_k)|\hcap_k^H\textbf{w}_k|^2 \geq \gamma_k\displaystyle\sum_{\ell=1}^{K} \lVert\textbf{T}_k\textbf{w}_{\ell}\lVert^2 + \gamma_k\sigma_k^2, \quad \forall k.
\end{equation} 
It is clear that the phases of beamformer vectors $\textbf{w}_k$ do not affect the objective and constraints expressions. Hence, we can express the SINR constraints as
\begin{equation} \label{GSB_eq3}
\sqrt{1+\gamma_k^{-1}}\text{Re}\{\hcap_k^H\textbf{w}_k\} \geq \sqrt{\displaystyle\sum_{\ell=1}^{K} \lVert\textbf{T}_k\textbf{w}_{\ell}\lVert^2 + \sigma_k^2}, \quad \forall k.
\end{equation} 
Per-RRH transmit power constraints can be written as
\begin{equation} \label{GSB_eq4}
\sqrt{\displaystyle\sum_{k=1}^K \lVert \textbf{B}_n\textbf{w}_k \lVert^2} \leq \sqrt{P_t} \quad \forall n.
\end{equation} 
Both (\ref{GSB_eq3}) and (\ref{GSB_eq4}) are second order cone constraints. Therefore, we can optimize the beamformers using a second order cone problem (P4). 
\begin{align} \label{P4}
\text{(P4)} \: &\min_{\textbf{w}_{k}} \: \displaystyle\sum_{k=1}^K\displaystyle\sum_{n=1}^N 2\sqrt{\epsilon_1\epsilon_2r_k}\lVert\textbf{B}_n\textbf{w}_{k}\lVert, \\
\text{s.t.} \: \: & \sqrt{1+\gamma_k^{-1}}\text{Re}\{\hcap_k^H\textbf{w}_k\} \geq \sqrt{\displaystyle\sum_{\ell=1}^{K} \lVert\textbf{T}_k\textbf{w}_{\ell}\lVert^2 + \sigma_k^2}, \quad \forall k, \notag \\
&\sqrt{\displaystyle\sum_{k=1}^K \lVert \textbf{B}_n\textbf{w}_k \lVert^2} \leq \sqrt{P_t}, \quad \forall n. \notag
\end{align}
We can use CVX \cite{CVX} software to solve second order cone programming (SOCP) problem (P4). In \cite{Wired-GreenCRAN}, \cite{Wired-Imperfect-CSI}, \cite{GSB}, this method is referred as group sparse beamforming (GSB). In the related reference papers, it is stated that this problem produces sparse solutions. 

To enhance the performance of the GSB method, the links are prioritized according to the solution of (P4). After this step, as in ILR method, links are removed according to the priority of the links until obtaining a feasible solution. As in \cite{Wired-GreenCRAN}, \cite{Wired-Imperfect-CSI}, \cite{GSB}, we use a priority function considering both beamformer and channel $\ell_2$ norms.  
\begin{equation} \label{GSB_eq5}
p(k, n) = \lVert\hcap_{kn}\lVert \cdot \lVert\textbf{w}_{kn}\lVert, \quad \forall k, n 
\end{equation}
where $p(k, n)$ is the priority of link $(k, n)$. The steps of the algorithm are given below.

\noindent\rule{0.5\textwidth}{0.8pt} 
\vspace{2mm}
\textbf{Algorithm} (Group Sparse Beamforming, GSB) \vspace{-4.5mm} \\
\noindent\rule{0.5\textwidth}{0.4pt}
\begin{itemize}[leftmargin=0.2in, label={}]
\item 1) Solve (P4) and evaluate the priority of each link.
\item 2) Starting from the full cooperation case, remove links according to the priorities and solve (P2) at each step until obtaining a feasible solution.  
\end{itemize}
\vspace{-3mm}
\noindent\rule{0.5\textwidth}{0.4pt} \\

\noindent \textbf{Method 3: Successive Convex Approximation (SCA)} \\
In this method, we approximate the $\ell_0$ norm using an upper bound found from the first order Taylor series \cite{SCA1}-\hspace{-0.1mm}\cite{SCA2}. To apply the approximation, we use the fact that 
\begin{equation} \label{SCA_eq1}
\lVert x \lVert_0 \approx f(x) = \dfrac{x}{x+\theta} \leq f(x_0)+f^{'}(x_0)(x-x_0)
\end{equation}
where $\theta>0$ is a small number, $x \geq 0$ is a real number, $x_0$ is a real number close to $x$, $f(x)=\dfrac{x}{x+\theta}$ is a concave function and $f^{'}$ denotes its first order derivative. Using the idea in (\ref{SCA_eq1}), we can approximate the term in $P$ related to $\ell_0$ norm as
\begin{equation} \label{SCA_eq2}
\displaystyle\sum_{n=1}^N (1-\delta[\lVert \textbf{w}_{kn}\lVert] \approx \displaystyle\sum_{n=1}^N \dfrac{\theta \lVert \textbf{w}_{kn} \lVert^2+\lVert \textbf{w}_{kn}^{(t)} \lVert^4}{(\lVert \textbf{w}_{kn}^{(t)} \lVert^2+\theta)^2}
\end{equation}
where the beamformers $\textbf{w}_{kn}^{(t)}$ are found from the previous iteration. The right-hand side of (\ref{SCA_eq2}) is convex with respect to beamformers and hence the optimization can be performed with successive convex approximations by omitting the rank-1 constraints. In the second part of Appendix E, we show that the optimal solution of the relaxed problem is always rank-1. The corresponding problem can be expressed as 
\begin{align} \label{P5}
\text{(P5)} \: &\min_{\textbf{W}_{k}} \: \displaystyle\sum_{k=1}^K\displaystyle\sum_{n=1}^N \Big(\dfrac{\epsilon_1r_k \theta \tr( \textbf{B}_n\textbf{W}_{k})+[\tr(\textbf{B}_n\textbf{W}_{k}^{(t)})]^2}{(\tr(\textbf{B}_n\textbf{W}_{k}^{(t)})+\theta)^2} \notag \\
&\qquad \qquad \qquad +\epsilon_2\tr(\textbf{B}_n\textbf{W}_{k})\Big), \\
&\displaystyle\sum_{k=1}^K \tr(\textbf{B}_n\textbf{W}_{k}) \leq P_t \quad \forall n=1, 2, \ldots, N, \notag \\
&\text{SINR}_k \geq \gamma_k, \quad \textbf{W}_k \succeq 0 \quad \forall k=1, 2, \ldots, K, \notag
\end{align}
where $\textbf{W}_{k}^{(t)}$'s are found from the previous iteration. The steps of the algorithm are given below.

\noindent\rule{0.5\textwidth}{0.8pt} 
\vspace{2mm}
\textbf{Algorithm} (Successive Convex Approximation, SCA) \vspace{-4.5mm} \\
\noindent\rule{0.5\textwidth}{0.4pt} \\
\textbf{Step 1, Initialization:} 
\begin{itemize}[leftmargin=0.2in, label={}]
\item Solve (P2) for full cooperation case. Set the initial values of $\textbf{W}_k^{(0)}$ for $k=1, 2, \ldots, K$. Define $\theta=10^{-5}, \: t_{\text{max, 2}}=10^3, \: \Delta_2=10^{-5}, \: \kappa_2 = 10^{-3}$.
\end{itemize}
\textbf{Step 2, Successive Convex Optimizations:} 
\begin{itemize}[leftmargin=0.2in, label={}]
\item For $t=1, 2, \ldots, t_{\text{max, 2}}$ repeat the following steps:
\item 1) Solve (P5) using $\textbf{W}_k^{(t-1)}$'s found in the previous iteration. 
\item 2) Check the condition for termination:
\begin{center} If $|P^{(t)}-P^{(t-1)}|/P^{(t)}<\Delta_2$, then terminate. \end{center}
\item After the termination, evaluate the $\ell_2$ norms of the final beamformer vectors. 
\end{itemize}
\textbf{Step 3, Find the link configuration:} 
\begin{itemize}[leftmargin=0.2in, label={}]
\item 1) Form a network link matrix $\textbf{C}$ by removing the links with small norms, i.e., $\left\Vert \textbf{w}_{kn} \right\Vert < \kappa_2 P_t$.
\item 2) Solve (P2) using $\textbf{C}$ with convex optimization. 
\end{itemize}
\vspace{-3mm}
\noindent\rule{0.5\textwidth}{0.4pt} \\

\section{Complexity Comparison}
In general, we can measure the computational complexity of an iterative method as the product of number of iterations and the complexity at each iteration. At each iteration of ILR, ES, MM, GSB, SCA, the main component of complexity is related to the convex optimization and all other operations can be neglected. For GSB method, the complexity of initial SOCP solution should also be evaluated. We use Self Dual Minimization (SeDuMi) \cite{SeDuMi} software (included in CVX \cite{CVX} package) as the convex optimization and SOCP solver. For the convex optimization based problems (P2), (P3), (P5), at each iteration, we minimize $c^Hx$ subject to $Ax=b$ where $x \in \mathbb{C}^{n}$ is the vector of all unknowns and $A \in \mathbb{C}^{m \times n}, \: b\in\mathbb{C}^m, \: c\in \mathbb{C}^{n}$ are known vectors/matrices. We know by \cite{SeDuMi} that the corresponding computational complexity is $\mathcal{O}(n^2m^{2.5}+m^{3.5})$ for SeDuMi. The corresponding $m$ and $n$ values are calculated as 
\begin{equation} \label{Complexity1}
\begin{aligned}
\text{(P2)} : \: \: & m=2K+N, \: \: n=KN^2L^2+K+N, \\
\text{(P3)} , \text{(P5)} : \: \: & m=K+N, \: \: n=KN^2L^2+K+N.
\end{aligned}
\end{equation}
Considering the number of convex optimizations required for ILR and ES, we obtain that
\begin{equation} \label{Complexity2}
\begin{aligned}
\text{C}_{\text{ILR}} &= \mathcal{O}\left((NK-K)C_0\right), \\
\text{C}_{\text{ES}} &= \mathcal{O}\left((2^N-1)^KC_0\right).
\end{aligned}
\end{equation}
where $\text{C}_{\text{ILR}}, \: \text{C}_{\text{ES}}$ are complexities of ILR and ES, respectively, and
\begin{equation} \label{C_0}
C_0=(2K+N)^{2.5}([KN^2L^2+K+N]^2+2K+N). 
\end{equation}
The complexities of MM and SCA can be expressed as
\begin{equation} \label{Complexity3}
\begin{aligned}
\text{C}_{\text{MM}}  &= \mathcal{O}\left(N_{\text{MM}}C_1\right), \\
\text{C}_{\text{SCA}} &= \mathcal{O}\left(N_{\text{SCA}}C_1\right),
\end{aligned}
\end{equation}
where $\text{C}_{\text{MM}}, \: \text{C}_{\text{SCA}}, \: N_{\text{MM}}, \: N_{\text{SCA}}$ are complexities and number of iterations of MM and SCA, respectively, and 
\begin{equation} \label{C_1}
C_1=(K+N)^{2.5}([KN^2L^2+K+N]^2+K+N). 
\end{equation}
According to \cite{SOCP_paper}, the worst-case complexity of a SOCP problem can be calculated as $\mathcal{O}(\sqrt{n_1}n_2^2 n_3)$ where $n_1$ is the total number of constraints, $n_2$ is the total number of real optimization variables, and $n_3$ is the total dimension of constraints. In (P4), the number of constraints is equal to $K+N$, the number of real optimization variables is equal to $2KNL$. Each SINR constraint has dimension $KNL+1$ and each per-RRH transmit power constraint has dimension $KL$. Hence, the total dimension of constraints is equal to $K(KNL+1)+KNL$. As a result, we obtain the complexity of (P4) as $\mathcal{O}(C_2)$ where
\begin{equation} \label{C_2}
C_2 = \sqrt{K+N}4(KNL)^2[K(KNL+1)+KNL].
\end{equation}
After the prioritization of links by solving (P4), GSB method also solves (P2) iteratively by link removals. Therefore, the complexity of GSB can be calculated as
\begin{equation} \label{Complexity4}
\text{C}_{\text{GSB}} = \mathcal{O}(C_2+(NK-K)C_0). 
\end{equation}
For SCFA, we can show that the main component of the complexity is related to calculation of gradient. The gradient calculation consists of four parts related to $P_{\text{CP}}, \: P_{\text{RRH}}$, per-RRH power transmit constraints and SINR constraints. It can be shown that the complexities for the first three parts are $\mathcal{O}(KNL)$ and the complexity of the part related to SINR constraints is $\mathcal{O}(K^2N^2L^2)$. Hence, we have 
\begin{equation} \label{Complexity5}
\text{C}_{\text{SCFA}} = \mathcal{O}\left(N_{\text{SCFA,1}}K^2N^2L^2 + N_{\text{SCFA,2}}C_0\right)
\end{equation}
where $\text{C}_{\text{SCFA}}$ is the complexity of SCFA, $N_{\text{SCFA,1}}$ is the number of iterations for gradient descent, and $N_{\text{SCFA,2}}$ is the number of convex optimizations in the Step 3 of the algorithm. Using (\ref{Complexity5}), (\ref{C_0}) and the fact that $(x+y)^{2.5} \geq 2^{2.5}(xy)^{1.25}$ for any $x, y>0$, \footnote{Here we use well-known Arithmetic-Geometric Mean Inequality to show that $x+y \geq 2\sqrt{xy}$.} it can be shown that 
\begin{align} \label{Complexity6}
\text{C}_{\text{SCFA}} &\approx \mathcal{O}(N_{\text{SCFA,1}}K^2N^2L^2 + N_{\text{SCFA,2}}K^{3.25}N^{5.25}L^4), \notag \\
\text{C}_{\text{ILR}}  &\approx \mathcal{O}(K^{4.25}N^{6.25}L^4), \\
\text{C}_{\text{ES}}  &\approx \mathcal{O}(2^{KN}K^{3.25}N^{5.25}L^4). \notag
\end{align}
Similarly, the complexities of MM, SCA and GSB can be approximately written as 
\begin{align} \label{Complexity7}
\text{C}_{\text{MM}}  &\approx \mathcal{O}(N_{\text{MM}} K^{3.25}N^{5.25}L^4), \notag \\
\text{C}_{\text{SCA}} &\approx \mathcal{O}(N_{\text{SCA}} K^{3.25}N^{5.25}L^4), \\
\text{C}_{\text{GSB}} &\approx \mathcal{O}(K^{4.25}N^{3.25}L^3 + K^{4.25}N^{6.25}L^4). \notag
\end{align}
In Table 1-2, we see the problems solved by each algorithm and the approximate complexities, respectively.

{\renewcommand{\arraystretch}{1.2}
\begin{table}[H]
\captionsetup{format=plain, labelsep=period, labelfont={ieeeblue,bf,small}, justification=justified}
\caption{Problems solved for each algorithm}
\begin{center}
\vspace{-4mm}
\begin{tabular}{| c | c |}
\hline
Method & Problems Solved \\
\hline
SCFA & Gradient Descent defined by (\ref{Q_2})-(\ref{step-size2}), (P2) \\
\hline
ILR & (P2) \\
\hline
ES &  (P2) \\
\hline
MM & (P3) \\
\hline
GSB & (P2), (P4) \\
\hline
SCA & (P5) \\
\hline
\end{tabular}
\end{center}
\end{table}
} 

\vspace{-5mm}

{\renewcommand{\arraystretch}{1.4}
\begin{table}[H]
\captionsetup{format=plain, labelsep=period, labelfont={ieeeblue,bf,small}, justification=justified}
\caption{Approximate complexities for algorithms}
\begin{center}
\vspace{-4mm}
\begin{tabular}{| c | c |}
\hline
Method & Approximate Complexity \\
\hline
SCFA & $\mathcal{O}(N_{\text{SCFA,1}}K^2N^2L^2 + N_{\text{SCFA,2}}K^{3.25}N^{5.25}L^4)$ \\
\hline
ILR & $\mathcal{O}(K^{4.25}N^{6.25}L^4)$ \\
\hline
ES &  $\mathcal{O}\left((2^{KN}K^{3.25}N^{5.25}L^4\right)$ \\
\hline
MM & $\mathcal{O}(N_{\text{MM}} K^{3.25}N^{5.25}L^4)$ \\
\hline
GSB &  $\mathcal{O}(K^{4.25}N^{3.25}L^3 + K^{4.25}N^{6.25}L^4)$ \\
\hline
SCA &  $\mathcal{O}(N_{\text{SCA}} K^{3.25}N^{5.25}L^4)$ \\
\hline
\end{tabular}
\end{center}
\end{table}
}

\vspace{-3mm}

The simulation results show that $N_{\text{SCFA,1}}<100, \: N_{\text{SCFA,2}}<10, \: N_{\text{MM}}<100, \: N_{\text{SCA}}<100$ holds for most of the cases. Considering this result, we conclude that for small $K, N, L$ values, the complexity of SCFA is higher compared to ILR, MM, SCA, and GSB. On the other hand, when $K, N, L$ are large, the complexity of SCFA becomes much smaller than those of ILR, MM, SCA, and GSB. In any case the optimal algorithm ES is the most complex one. \\

\section{Theoretical Bounds on $P$}
In this section, we find lower and upper bounds for the optimum value $P_{\text{opt}}$ of the total power spent. The main result is stated in Theorem 3. \\

\noindent \textbf{Theorem 3:} $P_{\ell} \leq P_{\text{opt}} \leq P_u$ where
\begin{align}
\label{P_u} P_u &= \epsilon_1N\displaystyle\sum_{k=1}^K\log_2(1+\gamma_k)+\epsilon_2NP_t \\
\label{P_l} P_{\ell} &= \displaystyle\sum_{k=1}^K \mathcal{P}_k \\
\mathcal{P}_k &= \min_{S_k}\epsilon_1 \log_2(1+\gamma_k)|S_k|+\epsilon_2\dfrac{\gamma_k\sigma_k^2}{\lambda_{\text{max}}\left(\hcap_k^{\prime}(\hcap_k^{\prime})^H-\gamma_k\textbf{D}_k^{\prime}\right)} \notag
\end{align}
$S_k=\{n_1, n_2, \ldots, n_r\}$ is the index set of RRHs serving the user $k$, 
\begin{equation} \label{P_l2}
\begin{aligned}
\hcap_k^{\prime} &= [\hcap_{n_1}^T \: \hcap_{n_2}^T \: \cdots \: \hcap_{n_r}^T]^T, \\
\textbf{D}_k^{\prime} &= \text{diag}\left(\sigma_{kn_1}^2\textbf{I}_L, \sigma_{kn_2}^2\textbf{I}_L, \ldots, \sigma_{kn_r}^2\textbf{I}_L\right). 
\end{aligned}
\end{equation}
To obtain a feasible solution, $\lambda_{\text{max}}\left(\hcap_k^{\prime}(\hcap_k^{\prime})^H-\gamma_k\textbf{D}_k^{\prime}\right)>0$ should satisfy for all $k$. We evaluate $\mathcal{P}_k$ by applying the minimization over all possible $S_k$ sets. \\

\noindent \textbf{Proof: } The proof is given in Appendix F. \\

\noindent Notice that if $P_{\ell}>P_u$ for some channel vectors, we can directly say that the problem is infeasible without doing any optimization. In such a case, as it is not possible to serve all users, we may try to find which users can be served. This is referred as the user admission problem \cite{Admission} in the literature and out of scope of this study.
\vspace{-2mm}
\section{Simulation Results}
In this section, we compare the performances of the algorithms under various cases. Throughout the simulations, we assume that $\gamma_k = \gamma, \: \sigma_k=\sigma$ and $r_k=\log_2\left(1+\gamma\right) \quad \forall k$. We use a realistic channel model including path-loss, shadowing and small-scale fading defined in a 3GPP standard \cite{3GPP}. We consider a circular region in which RRHs and UEs are distributed uniformly.\footnote{To avoid channel model inconsistencies, configurations where RRH-to-UE distances are all at least 50 meters are considered.} In Table 3, the model parameters are presented. 

{\renewcommand{\arraystretch}{1.2}
\begin{table}[H]
\captionsetup{format=plain, labelsep=period, labelfont={ieeeblue,bf,small}, justification=justified}
\caption{Model parameters used in simulations}
\begin{center}
\vspace{-4mm}
\begin{tabular}{| c | c |}
\hline
Cell radius & $0.5$ km \\
\hline
Path-loss ($P_L$) & $P_L=128.1+37.6\log_{10}d$ where $d$ is in km \\
\hline
RRH/UE antenna gain & 0 dBi \\
\hline
Shadowing model/variance & Log-normal, 10 dB \\
\hline
UE Noise Figure (NF) & 9 dB \\
\hline
Bandwidth (BW) & 10 MHz \\
\hline
$\sigma^2$ & $-174+10\log_{10}\text{BW}+\text{NF} = -95$ dBm \\
\hline
Small-scale fading model & Rayleigh \\
\hline
$\epsilon_1$ & $5 \text{W} \cdot \text{Hz}/\text{bps}$ \\
\hline
$\epsilon_2$ & 2 \\
\hline
\end{tabular}
\end{center}
\end{table}
}

\vspace{-5mm}

In this study, we fix the parameters $\epsilon_1, \epsilon_2$ which are related to infrastructure and hardware quality of the network. We also assume that all user equipments are identical ($\sigma_k=\sigma, \: \forall k$) and priority of users are equal ($\gamma_k=\gamma, \: \forall k$). All algorithms analyzed can be operated with any combinations of these parameters. One can use a different parameter set to see the corresponding algorithm performances. 

To generate channel estimates and channel estimation errors, we assume that pilot signal powers are adjusted according to the channel amplitudes so that the power ratios of $\mathbb{E}(|\Delta\textbf{h}_{kn}|^2)/|\textbf{h}_{kn}|^2, \: \forall n, k$ are all equal to some known constant $\gamma_{\text{ch}}$. Here $\gamma_{\text{ch}}$ is a measure of channel estimation quality. Using the channel estimates and $\gamma_{\text{ch}}$, one can evaluate $\sigma_{kn}^2, \: \forall n, k$ accordingly.

In simulations, we observe the effect of parameters $\gamma, K, N, L, \gamma_{\text{ch}}, P_t$. We generate channel vectors so that the problem is feasible (it is possible to satisfy all constraints of the problem).\footnote{The feasibility of the problem (P1) is equivalent to feasibility of (P2) for full cooperation case.} We run $1000$ Monte-Carlo trials in each case. To make a comparison, we run ES method when $K, N, L$ are small. When these parameters are large ES will not be depicted due to impractical run-times.

\subsection{Effect of SINR threshold}
We know that the total power spent is an increasing function of SINR threshold $\gamma$. This fact can also be verified considering the lower bound expression given in Theorem 3. In Fig. 2-3, we see the performances of the algorithms as $\gamma$ varies. We take $P_t = 10 \text{W}, \gamma_{\text{ch}}=0.01$ in both cases and consider the results for triples $(K, N, L)=(3, 3, 2), (8, 4, 8)$. The results show that SCFA outperforms other algorithms (except for ES which is the optimal one) for all $\gamma$ values. 

\begin{figure}[ht]
\centering
\captionsetup{format=plain, labelsep=period, labelfont={ieeeblue,bf,small}, justification=justified}
\includegraphics[width=\figwidth\textwidth]{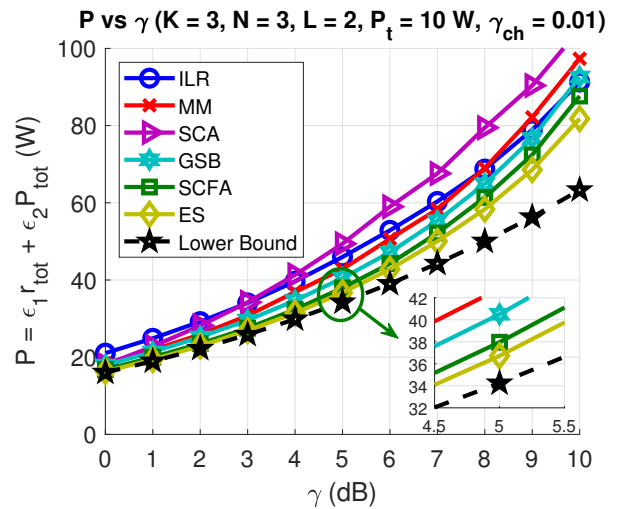}
\caption{$P$ vs $\gamma$. $K=3, N=3, L=2, P_t = 10 \text{W}, \gamma_{\text{ch}}=0.01$.}
\end{figure}

In Fig. 2, we observe that ES requires slightly less power than SCFA. The power difference between SCFA and ES is about 1 W for $\gamma=5$ dB. We also note that the performance of SCFA is very close to the theoretical lower bound for small $\gamma$ values. The power difference between SCFA and the bound is roughly 10 percent for $\gamma=5$ dB. As $\gamma$ increases, the lower bound becomes too optimistic. This is due to the fact that the lower bound can be achieved when the inter-user interference is perfectly eliminated which becomes harder for large SINR thresholds. 

\begin{figure}[ht]
\centering
\captionsetup{format=plain, labelsep=period, labelfont={ieeeblue,bf,small}, justification=justified}
\includegraphics[width=\figwidth\textwidth]{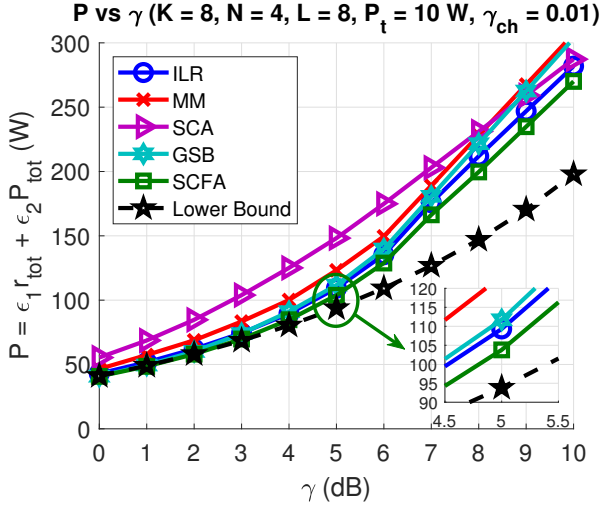}
\caption{$P$ vs $\gamma$. $K=8, N=4, L=8, P_t = 10 \text{W}, \gamma_{\text{ch}}=0.01$.}
\end{figure}

In Fig. 3, we also see that SCFA outperforms other methods and its performance is close to the lower bound at small SINR thresholds. Considering the results, we can say that the performance of SCFA is satisfactory even when there is imperfect channel knowledge.

\subsection{Effect of Number of Users} 
As the number of users increases, we expect higher power consumption to satisfy all SINR constraints. 

\begin{figure}[ht]
\centering
\captionsetup{format=plain, labelsep=period, labelfont={ieeeblue,bf,small}, justification=justified}
\includegraphics[width=\figwidth\textwidth]{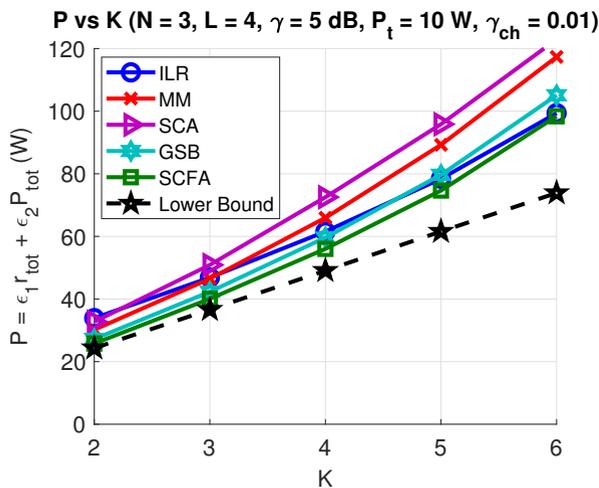}
\caption{$P$ vs $K$. $N=3, L=4, \gamma=5 \: \text{dB},  P_t = 10 \text{W}, \gamma_{\text{ch}}=0.01$.}
\end{figure}

In Fig. 4, we observe the effect of number of users. We see that SCFA outperforms other methods for all $K$ values. The results show that the power difference between SCFA and the bound is very low for small $K$ values. The difference becomes large as $K$ increases. The reason is similar to that of the large $\gamma$ case. To achieve the bound, the interference due to other users should be completely eliminated. As $K$ increases it becomes harder to eliminate this interference and hence the power differences between the methods and the bound become larger.

\subsection{Effect of Number of RRHs and Number of RRH Antennas} 
When the number of RRHs and/or the number of RRH antennas increases, the dimension of the augmented beamformer vector $\textbf{w}$ becomes higher which enables a better resource allocation and provides a lower power consumption. 

\begin{figure}[H]
\centering
\captionsetup{format=plain, labelsep=period, labelfont={ieeeblue,bf,small}, justification=justified}
\includegraphics[width=\figwidth\textwidth]{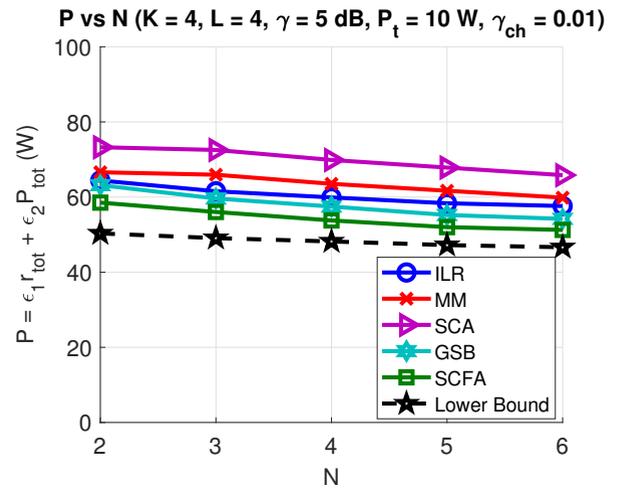}
\caption{$P$ vs $N$. $K=4, L=4, \gamma=5 \: \text{dB}, P_t = 10 \text{W}, \gamma_{\text{ch}}=0.01$.}
\end{figure}

\begin{figure}[H]
\centering
\captionsetup{format=plain, labelsep=period, labelfont={ieeeblue,bf,small}, justification=justified}
\includegraphics[width=0.47\textwidth]{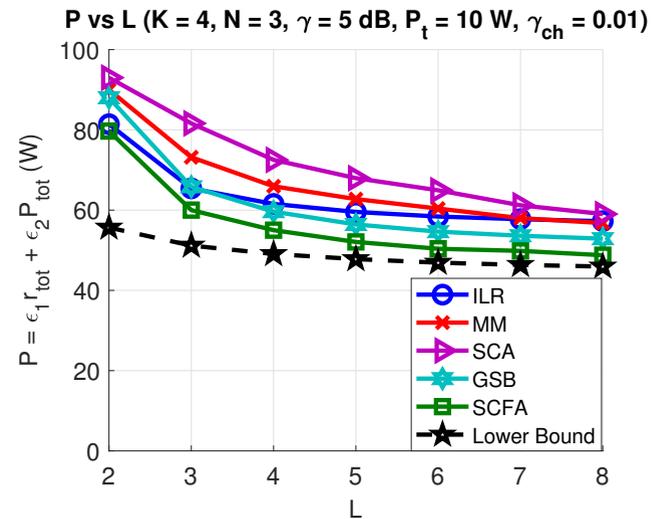}
\caption{$P$ vs $L$. $K=4, N=3, \gamma=5 \: \text{dB}, P_t = 10 \text{W}, \gamma_{\text{ch}}=0.01$.}
\end{figure}

Fig. 5-6 show the effect of number of RRHs and number of RRH antennas. As in the previous cases, SCFA performance is better compared to other methods for all $N$ and $L$ values. We observe that the power values of all methods are lower bounded as $N$ or $L$ increases. This is due to the fact that for a fixed $K$, optimal strategy uses some set of links between RRHs and UEs, and if there are sufficient links in the system, adding more links does not improve the performance much. We also observe that for large $N$ or $L$, the performance of SCFA becomes very close to the bound. This shows that the bound is very tight and SCFA performance is near optimal at these values.

\subsection{Effect of Channel Estimation Quality}
We know by (\ref{SINR_2}) that a larger channel estimation error corresponds a lower SINR which results in a higher power consumption. 

\begin{figure}[ht]
\centering
\captionsetup{format=plain, labelsep=period, labelfont={ieeeblue,bf,small}, justification=justified}
\includegraphics[width=0.47\textwidth]{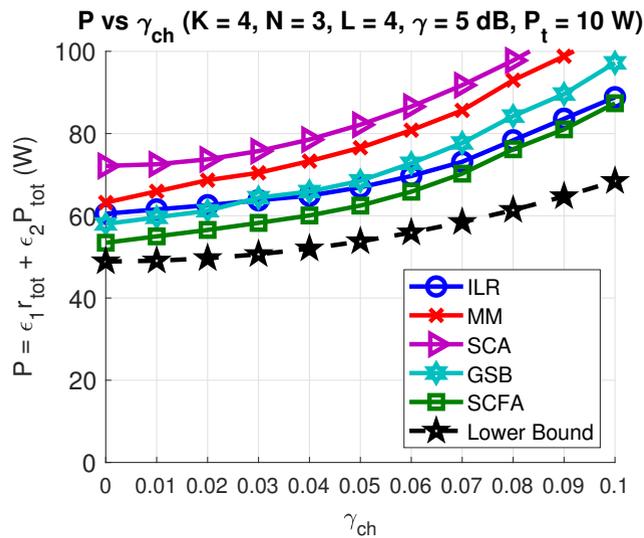}
\caption{$P$ vs $\gamma_{\text{ch}}$. $K=4, N=3, L=4, \gamma=5 \: \text{dB}, P_t = 10 \text{W}$.}
\end{figure}

\begin{figure}[ht]
\centering
\captionsetup{format=plain, labelsep=period, labelfont={ieeeblue,bf,small}, justification=justified}
\includegraphics[width=0.47\textwidth]{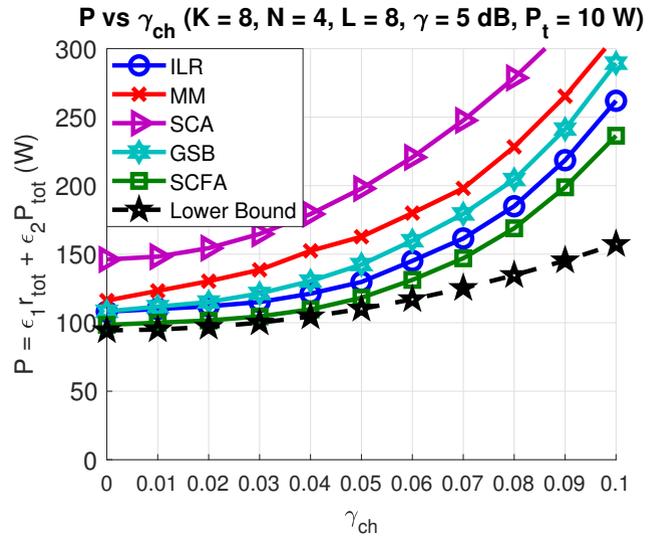}
\caption{$P$ vs $\gamma_{\text{ch}}$. $K=8, N=4, L=8, \gamma=5 \: \text{dB}, P_t = 10 \text{W}$.}
\vspace{-4mm}
\end{figure}

Fig. 7-8 present the effect of channel estimation quality. We see that when the channel estimation quality is poor, the power required to satisfy the constraints of the problem becomes large. In $(K, N, L)=(4, 3, 4)$ case, for all methods, there is a roughly $50$ percent increase in the power between $\gamma_{\text{ch}}=0$ (perfect CSI) and $\gamma_{\text{ch}}=0.1$. The increase in the power becomes larger than $100$ percent for $(K, N, L)=(8, 4, 8)$ case for all methods. This shows that channel estimation quality has a significant effect on the power consumption. On the other hand, by means of the robustness of the algorithm designs, increasing the channel estimation error variance by a factor of 10 yields only 2 times higher power consumption. We also observe that the performances of all methods are far away from the bound when the channel estimation quality is not good enough. This shows that inter-user interference cannot be completely eliminated when the channel estimation quality is poor. 

\subsection{Effect of per-RRH Power Transmit Constraint}
RRH power transmit constraint is one of the limiting factors of the performance. The lower bound derivation does not include the effect of power transmit constraints and hence we expect a constant lower bound as $P_t$ varies. On the other hand, when $P_t$ is small, it becomes a limiting factor and hence it affects the performance of the algorithms. 

\begin{figure}[ht]
\centering
\captionsetup{format=plain, labelsep=period, labelfont={ieeeblue,bf,small}, justification=justified}
\includegraphics[width=0.47\textwidth]{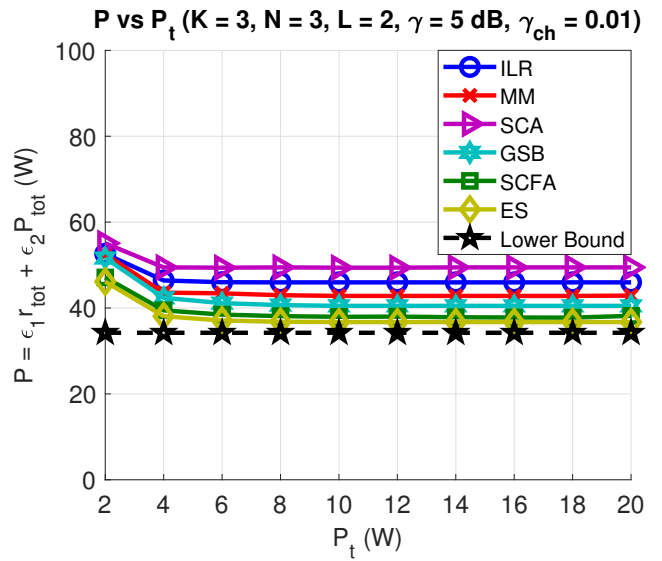}
\caption{$P$ vs $P_t$. $K=3, N=3, L=2, \gamma=5 \: \text{dB}, \gamma_{\text{ch}}=0.01$.}
\end{figure}

Fig. 9-10 show the effect of per-RRH transmit power constraint. There exists a value for $P_t$ after which the performances of all methods remain constant. After that value, the power spent by each RRH becomes low enough to satisfy transmit power constraints, hence further increasing $P_t$ does not affect the performance. 

\begin{figure}[ht]
\centering
\captionsetup{format=plain, labelsep=period, labelfont={ieeeblue,bf,small}, justification=justified}
\includegraphics[width=0.45\textwidth]{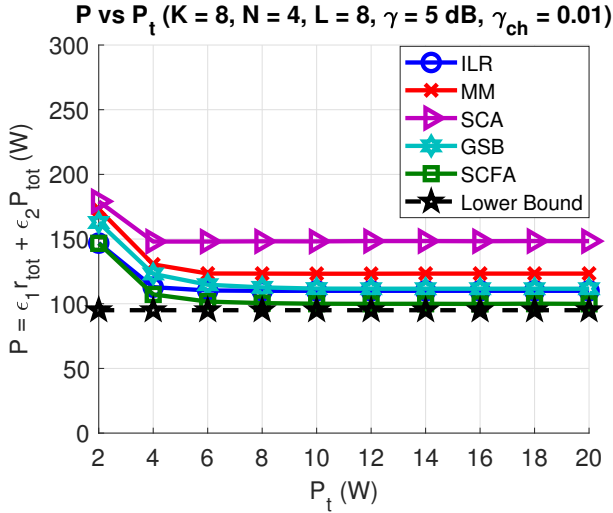}
\caption{$P$ vs $P_t$. $K=8, N=4, L=8, \gamma=5 \: \text{dB}, \gamma_{\text{ch}}=0.01$.}
\vspace{-4mm}
\end{figure}

\subsection{Average Run-Time Comparison}
In Section IV, we evaluated the approximate complexity values of each algorithm. In this part, we aim to verify the results by measuring the average run-times of the algorithms.  

\begin{figure}[ht]
\centering
\captionsetup{format=plain, labelsep=period, labelfont={ieeeblue,bf,small}, justification=justified}
\includegraphics[width=0.38\textwidth]{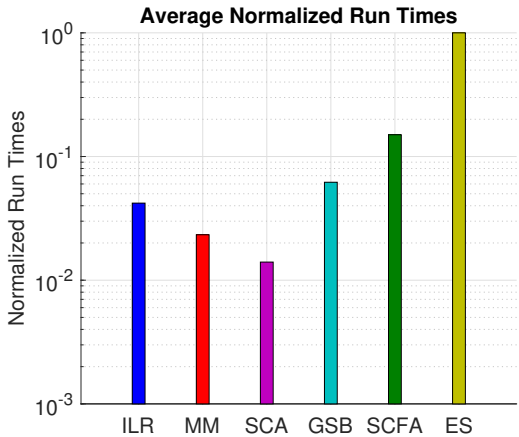}
\caption{Average Complexity Comparison $(K, N, L)=(3, 3, 2)$.}
\vspace{-4mm}
\end{figure}

Fig. 11-12 present the average normalized\footnote{We normalize the run-times so that the maximum run-time is equal to 1.} run-times of each method for $(K, N, L)=(3, 3, 2)$ and $(K, N, L)=(8, 4, 8)$ where we calculate the run-times by averaging the results over all previously described simulations with given $(K, N, L)$ triples.\footnote{We use Matlab 2020b to run all algorithms.} We see that although the performance of SCFA is the best, its computational complexity is higher compared to all methods other than ES for $(K, N, L)=(3, 3, 2)$ case. Nevertheless, its complexity is much lower than that of the optimal algorithm ES. On the other hand, the complexity of SCFA becomes much less than those of ILR, MM, SCA and GSB for $(K, N, L)=(8, 4, 8)$ case.\footnote{We cannot run ES for large $K, N, L$ due to its high computational complexity.} This fact is also verified in Complexity Comparison section.

\begin{figure}[ht]
\centering
\captionsetup{format=plain, labelsep=period, labelfont={ieeeblue,bf,small}, justification=justified}
\includegraphics[width=0.38\textwidth]{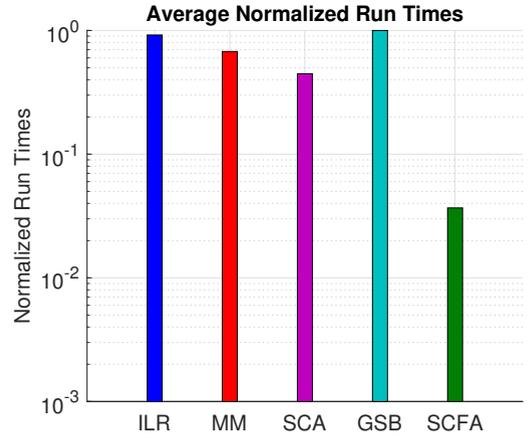}
\caption{Average Complexity Comparison $(K, N, L)=(8, 4, 8)$.}
\vspace{-4mm}
\end{figure}

\section{Conclusions}
In this study, we analyze the robust beamforming design problem for downlink C-RAN with imperfect channel state information. We consider both the fronthaul power which is related to fronthaul data transfer from CP to RRHs, and total power spent in RRHs to transmit user data to UEs. The total power spent in the C-RAN cluster is minimized under user SINR and per-RRH power transmit constraints. The original problem includes discontinuous terms making the optimization challenging. To do a proper beamforming design, we transform the problem into a smooth constraint-free function minimization problem and optimization is performed using gradient descent iteratively. The initial point of the algorithm is found by the solution of the full cooperation case which can be optimally solved by convex optimization. As a comparison, we also consider a heuristic search method applying successive convex optimizations to the equivalent combinatorial convex version of the original problem, and sparse beamforming techniques applying $\ell_0$ norm approximations. To make optimality analysis, we find a novel theoretical performance bound.

The detailed simulations show that the proposed method SCFA provides an efficient solution. Its performance is better than those of other non-optimal algorithms analyzed, and the power values found for SCFA is close to the lower bound and to the optimal but impractical algorithm ES. For some cases (small $K$ or large $N$ or large $L$), the power values of SCFA and the bound becomes very close. This proves that the bound that we derived is very tight and SCFA performance is near optimal at those values of network parameters. 

Throughout the paper, we assume a wired fronthaul to send user data from CP to RRHs. As a future study, one may consider a wireless fronthaul case where RRHs apply some relaying strategy to forward the user data to UEs. The smooth constraint-free approximation idea proposed in this study can be modified accordingly to solve the joint fronthaul and access link beamforming problem in wireless fronthaul case.   


\section*{Appendix A (Achievability of Rate)}
We use the idea given in \cite{InformationTheory} to show that the rate $\log_2(1+\text{SINR}_k)$ is achievable for $k$-th user where $\text{SINR}_k$ is defined by (\ref{SINR_2}). We find a lower bound to the mutual information $I(r_k; s_k)$ between the received signal $r_k$ and the information signal $s_k$. Assuming that $s_k$ is zero-mean and Gaussian distributed, for any given complex constant $\alpha$ we can show that 
\begin{align} 
I(r_k; s_k)&=h(s_k)-h(s_k | r_k) \notag \\
\label{App1_1} &= h(s_k)-h(s_k-\alpha r_k | r_k) \\
\label{App1_2} &\geq h(s_k)-h(s_k-\alpha r_k) \\
\label{App1_3} &\geq \log\left(\pi e \mathbb{E}\left[|s_k|^2\right]\right) - \log\left(\pi e \mathbb{E}\left[|s_k-\alpha r_k|^2\right]\right) \\
\label{App1_4} &= \log\left(\dfrac{\mathbb{E}\left[|s_k|^2\right]}{\mathbb{E}\left[|s_k-\alpha r_k|^2\right]}\right).
\end{align}
(\ref{App1_1}) is obtained using the fact that the entropy $h(\cdot)$ is invariant under translation. (\ref{App1_2}) can be shown using the fact that conditioning decreases entropy. As $s_k$ is zero-mean and Gaussian, we get $h(s_k) = \log\left(\pi e \mathbb{E}\left[|s_k|^2\right]\right)$. Using the facts that the entropy is maximized for Gaussian distribution when the variance is fixed and $r_k$ is also zero-mean, we have $h(s_k-\alpha r_k) \leq  \log\left(\pi e \mathbb{E}\left[|s_k-\alpha r_k|^2\right]\right)$. Therefore, we obtain the result in (\ref{App1_3}). Finally, using some simple calculations for log terms, we get the final result in (\ref{App1_4}). 

(\ref{App1_1})-(\ref{App1_4}) are true for any $\alpha$ and specifically we choose $\alpha=\mathbb{E}\left[r_k^{*}s_k\right]/\mathbb{E}\left[|r_k|^2\right]$ to get
\begin{equation} \label{App2}
I(r_k; s_k) \geq \log\left(1 + \dfrac{|\mathbb{E}\left[r_k^{*}s_k\right]|^2}{\mathbb{E}\left[|r_k|^2\right] \cdot \mathbb{E}\left[|s_k|^2\right] - |\mathbb{E}\left[r_k^{*}s_k\right]|^2}\right).
\end{equation}
Using the equation of $r_k$ in (\ref{r_k_1}) and the fact $\mathbb{E}\left[|s_k|^2\right]=1$, we obtain that $|\mathbb{E}\left[r_k^{*}s_k\right]|^2 = P_d$ and $\mathbb{E}\left[|r_k|^2\right] = P_d + P_{m}+P_{\text{int}}+P_n$ where $P_d, P_{m}, P_{\text{int}}, P_n$ are defined in (\ref{P_d}). Therefore, we conclude that the mutual information between $r_k$ and $s_k$ is at least $\log_2(1+\text{SINR}_k)$ bits where $\text{SINR}_k = \dfrac{P_d}{P_{m}+P_{\text{int}}+P_n}$.

\section*{Appendix B (Proof of Rank-1 Solution of (P2))}
Using the idea in \cite{SCA1}, we can prove that the relaxed version of (P2) has always rank-1 solution. Firstly, we define the Lagrangian function of the problem $\mathcal{L}$ as 
\begin{align}
&\mathcal{L}=\displaystyle\sum_{k=1}^K \tr(\textbf{W}_k)-\displaystyle\sum_{k=1}^K\lambda_k\rho_k+\displaystyle\sum_{k=1}^K\mu_k\displaystyle\sum_{n=1}^N c_{kn}\tr(\textbf{B}_n\textbf{W}_k) \notag \\ 
&+\displaystyle\sum_{n=1}^N v_n\left(\displaystyle\sum_{k=1}^K\tr(\textbf{B}_n\textbf{W}_k)-P_t\right)-\displaystyle\sum_{k=1}^K\tr(\textbf{W}_k\textbf{Z}_k)
\end{align}
where $\{\lambda_k\}_{k=1}^K, \{\mu_k\}_{k=1}^K, \{v_n\}_{n=1}^N$ are non-negative Lagrange multipliers, $\rho_k$ is defined in (\ref{rho}), $\textbf{Z}_k$ is the dual positive semi-definite matrix associated with the positive semi-definite optimization variable $\textbf{W}_k$. For the optimal solution of the convex problem (P2), the condition $\textbf{W}_k \textbf{Z}_k =\textbf{0}$ should satisfy for all $k$. Furthermore, by the first order optimality condition, we also have
\begin{align}
\dfrac{\partial \mathcal{L}}{\partial \textbf{W}_k}&=\textbf{I}-\lambda_k(1+\gamma_k)\Hcap_k+\displaystyle\sum_{\ell=1}^K\lambda_{\ell}\gamma_{\ell}\Htilde_{\ell} \notag \\
&+\displaystyle\sum_{n=1}^N(\mu_k c_{kn}+v_n)\textbf{B}_n-\textbf{Z}_k \\
&= \textbf{0}. \notag
\end{align}
It follows that 
\begin{equation}
\textbf{Z}_k = \textbf{I} + \textbf{F}_k - \lambda_k(1+\gamma_k)\hcap_k\hcap_k^H
\end{equation}
where 
\begin{equation} \textbf{F}_k=\displaystyle\sum_{\ell=1}^K\lambda_{\ell}\gamma_{\ell}\Htilde_{\ell}+\displaystyle\sum_{n=1}^N(\mu_kc_{kn}+v_n)\textbf{B}_n
\end{equation} 
is a positive semi-definite matrix. Therefore, the matrix $\textbf{I} + \textbf{F}_k$ is full-rank. The rank of $\lambda_k(1+\gamma_k)\hcap_k\hcap_k^H$ is at most $1$ and hence the rank of $\textbf{Z}_k$ is at least $NL-1$ as its dimension is $NL$. Using the equation $\textbf{W}_k \textbf{Z}_k =\textbf{0}$ and the rank inequality for matrix multiplication, we get 
\begin{align}
\text{rank}(\textbf{W}_k) &\leq NL + \text{rank}(\textbf{W}_k \textbf{Z}_k) - \text{rank}(\textbf{Z}_k) \notag \\
&= NL - \text{rank}(\textbf{Z}_k) \\
&\leq 1. \notag
\end{align}
Assuming positive SINR threshold values, the zero beamformer cannot be a solution and hence we conclude that $\text{rank}(\textbf{W}_k)=1, \: \forall k$. 

\section*{Appendix C (Proof of Theorem 1)}
Assume on the contrary that there exists a user $k$ with $\text{SINR}_k > \gamma_k$. Choose a constant $0<\lambda<1$ and replace $\textbf{w}_k$ by $\lambda \textbf{w}_k$. In that case $\text{SINR}_{\ell}$ becomes larger for all $\ell \neq k$. If $\text{SINR}_k$ becomes also larger then all SINR constraints are satisfied. If $\text{SINR}_k$ becomes smaller, we can choose $\lambda$ small enough to ensure that in the new case $\text{SINR}_k \geq \gamma_k$ holds. This is due to the fact that the expression of $\text{SINR}_k$ is continuous with respect to $\lambda$. The original SINR value of the $k$-th user $\text{SINR}_{k,0}$ (corresponding to $\lambda_0=1$) is larger than $\gamma_k$ and if a new value $\text{SINR}_{k,1}<\text{SINR}_{k,0}$ is obtained for some $0<\lambda_1<1$, then by Intermediate Value Theorem, there exists a $\lambda_2$ with $0<\lambda_1<\lambda_2<1$ such that the corresponding SINR value $\text{SINR}_{k,2}$ satisfies $\gamma_k \leq \text{SINR}_{k,2}<\text{SINR}_{k,0}$. In both cases, the term $P_{\text{CP}}$ remains the same but $P_{\text{tot}}$ becomes smaller making $P$ smaller, which is a contradiction. Therefore, in the optimal case, $\text{SINR}_k = \gamma_k$ must hold for all $k$. 

\section*{Appendix D (Proof of Theorem 2)}
Notice that there are $2^{NK}$ possible $\textbf{C}$ matrices, however, in order to serve all users, each column of $\textbf{C}$ should include at least one $0$, making the total number smaller. For each column of $\textbf{C}$, there are $2^N-1$ possible arrangements (all possible arrangements except all $1$ case) and hence there are $(2^N-1)^K$ many possible $\textbf{C}$ matrices.

\section*{Appendix E (Proof of Rank-1 Solution of (P3) and (P5))}
We use the idea presented in Appendix B to show that the solutions of (P3) and (P5) are always rank-1. For (P3), we consider the Lagrangian function
\begin{align}
\mathcal{L}_1 &= \displaystyle\sum_{k=1}^K \left(r_k \displaystyle\sum_{n=1}^N \dfrac{\epsilon_1 c_{\theta} \tr(\textbf{B}_n\textbf{W}_k)}{\tr(\textbf{B}_n\textbf{W}_k^{(t)})+\theta} + \epsilon_2\tr(\textbf{W}_k)\right) \notag \\
&-\displaystyle\sum_{k=1}^K\alpha_{k,1}\rho_k+\displaystyle\sum_{n=1}^N \beta_{n,1}\left(\displaystyle\sum_{k=1}^K\tr(\textbf{B}_n\textbf{W}_k)-P_t\right)\\
&-\displaystyle\sum_{k=1}^K\tr(\textbf{W}_k\textbf{Z}_{k,1}) \notag
\end{align}
where $\{\alpha_{k,1}\}_{k=1}^K, \{\beta_{n,1}\}_{n=1}^N$ are non-negative Lagrange multipliers, $\rho_k$ is defined in (\ref{rho}), $\textbf{Z}_{k,1}$ is the dual positive semi-definite matrix associated with the positive semi-definite optimization variable $\textbf{W}_k$. For the optimal solution of the convex problem (P3), the condition $\textbf{W}_k \textbf{Z}_{k,1} =\textbf{0}$ should be satisfied for all $k$. By the first order optimality condition, we get
\begin{align}
\dfrac{\partial \mathcal{L}_1}{\partial \textbf{W}_k}&=r_k\displaystyle\sum_{n=1}^N\left(\dfrac{\epsilon_1c_{\theta}}{\tr(\textbf{B}_n\textbf{W}_k^{(t)})+\theta} + \beta_{n,1}\right)\textbf{B}_n + \epsilon_2\textbf{I} \notag \\
&+ \displaystyle\sum_{\ell=1}^K\alpha_{\ell,1}\gamma_{\ell}\Htilde_{\ell} - \alpha_{k,1}(1+\gamma_k)\Hcap_k - \textbf{Z}_{k,1}\\
&= \textbf{0}. \notag
\end{align}
It follows that 
\begin{equation}
\textbf{Z}_{k,1} = \epsilon_2\textbf{I} + \textbf{F}_{k,1} - \alpha_{k,1}(1+\gamma_k)\hcap_k\hcap_k^H
\end{equation}
where 
\begin{equation} 
\begin{aligned}
\textbf{F}_{k,1}&=\displaystyle\sum_{n=1}^N\left(\dfrac{r_k\epsilon_1c_{\theta}}{\tr(\textbf{B}_n\textbf{W}_k^{(t)})+\theta} + \beta_{n,1}\right)\textbf{B}_n\\
&+ \displaystyle\sum_{\ell=1}^K\alpha_{\ell,1}\gamma_{\ell}\Htilde_{\ell}.
\end{aligned}
\end{equation}
$\textbf{F}_{k,1}$ is a positive semi-definite matrix, and given that $\epsilon_2>0$, we have $\text{rank}(\epsilon_2\textbf{I})=NL$ and $\text{rank}(\alpha_{k,1}(1+\gamma_k)\hcap_k\hcap_k^H) \leq 1$. Therefore, we get $\text{rank}(\textbf{Z}_{k,1})\geq NL-1$. Using the equation $\textbf{W}_k \textbf{Z}_{k,1} =\textbf{0}$ and the rank inequality for matrix multiplication, we obtain the final result as 
\begin{align}
\text{rank}(\textbf{W}_k) &\leq NL + \text{rank}(\textbf{W}_k \textbf{Z}_{k,1}) - \text{rank}(\textbf{Z}_{k,1}) \notag \\
&= NL - \text{rank}(\textbf{Z}_{k,1}) \\
&\leq 1. \notag
\end{align}
The zero beamformer is not a solution for $\gamma_k>0$, and hence we obtain that $\text{rank}(\textbf{W}_k)=1, \: \forall k$. \\
For (P5), we again consider the Lagrangian $\mathcal{L}_2$, which can be written as
\begin{align}
&\mathcal{L}_2 = \displaystyle\sum_{k=1}^K \displaystyle\sum_{n=1}^N \left(\dfrac{\epsilon_1 r_k \tr(\textbf{B}_n\textbf{W}_k) + [\tr(\textbf{B}_n\textbf{W}_k^{(t)})]^2}{(\tr(\textbf{B}_n\textbf{W}_k^{(t)})+\theta)^2}\right) \notag \\
&+ \displaystyle\sum_{k=1}^K\epsilon_2\tr(\textbf{W}_k) - \displaystyle\sum_{k=1}^K\alpha_{k,2}\rho_k \\
&+\displaystyle\sum_{n=1}^N \beta_{n,2}\left(\displaystyle\sum_{k=1}^K\tr(\textbf{B}_n\textbf{W}_k)-P_t\right)-\displaystyle\sum_{k=1}^K\tr(\textbf{W}_k\textbf{Z}_{k,2}) \notag
\end{align}
where $\{\alpha_{k,2}\}_{k=1}^K, \{\beta_{n,2}\}_{n=1}^N$ are non-negative Lagrange multipliers, $\textbf{Z}_{k,2}$ is the dual positive semi-definite matrix associated with the positive semi-definite optimization variable $\textbf{W}_k$. For the optimal solution of the convex problem (P3), the condition $\textbf{W}_k \textbf{Z}_{k,2} =\textbf{0}$ should be satisfied for all $k$. By the first order optimality condition, we get
\begin{align}
\dfrac{\partial \mathcal{L}_2}{\partial \textbf{W}_k}&=\displaystyle\sum_{n=1}^N\left(\dfrac{\epsilon_1r_k}{(\tr(\textbf{B}_n\textbf{W}_k^{(t)})+\theta)^2} + \beta_{n,2}\right)\textbf{B}_n + \epsilon_2\textbf{I}  \notag \\
&+\displaystyle\sum_{\ell=1}^K\alpha_{\ell,2}\gamma_{\ell}\Htilde_{\ell} - \alpha_{k,2}(1+\gamma_k)\Hcap_k- \textbf{Z}_{k,2} \\
&= \textbf{0}. \notag
\end{align}
Hence, we get 
\begin{equation}
\textbf{Z}_{k,2} = \epsilon_2\textbf{I} + \textbf{F}_{k,2} - \alpha_{k,2}(1+\gamma_k)\hcap_k\hcap_k^H
\end{equation}
where 
\begin{equation}
\begin{aligned} 
\textbf{F}_{k,2}&=\displaystyle\sum_{n=1}^N\left(\dfrac{\epsilon_1r_k}{(\tr(\textbf{B}_n\textbf{W}_k^{(t)})+\theta)^2}+\beta_{n,2}\right)\textbf{B}_n  \\
&+ \displaystyle\sum_{\ell=1}^K\alpha_{\ell,2}\gamma_{\ell}\Htilde_{\ell}
\end{aligned}
\end{equation}
which is a positive semi-definite matrix. For $\epsilon_2>0$, the matrix $\epsilon_2\textbf{I}$ is full-rank and since the rank of $\alpha_{k,2}(1+\gamma_k)\hcap_k\hcap_k^H$ is at most $1$, we conclude that $\text{rank}(\textbf{Z}_{k,2}) \geq NL-1$. As in the previous part, we obtain that $\text{rank}(\textbf{W}_k)=1, \: \forall k$.

\section*{Appendix F (Proof of Theorem 3)}
For the upper bound, we use the fact that maximum fronthaul power is reached for full cooperation case and each RRH can transmit at most $P_t$ power. Hence, we obtain the result in (\ref{P_u}). For the lower bound, using the user SINR expressions given in (\ref{SINR_2}) and the fact 
\begin{equation} \label{App3}
\tr(\Htilde_k \textbf{W}_{\ell}) \geq 0, \: \forall k, \ell
\end{equation}
we get 
\begin{equation} \label{App4}
\tr(\Hcap_k\textbf{W}_k) \geq \gamma_k\tr(\textbf{D}_k\textbf{W}_k) + \gamma_k\sigma_k^2.
\end{equation}
Hence, we have
\begin{equation} \label{App5}
\textbf{w}_k^H\left(\hcap_k\hcap_k^H-\gamma_k\textbf{D}_k\right)\textbf{w}_k \geq \gamma_k\sigma_k^2.
\end{equation}
Since $S_k=\{n_1, n_2, \ldots, n_r\}$ is the index set of RRHs serving the user $k$, we have $\textbf{w}_{kn}=\textbf{0}$ for all $n \not\in S_k$. Therefore, we can write
\begin{equation} \label{App6}
\textbf{w}_k^H\left(\hcap_k\hcap_k^H-\gamma_k\textbf{D}_k\right)\textbf{w}_k = (\textbf{w}_k^{\prime})^H\left(\hcap_k^{\prime}(\hcap_k^{\prime})^H-\gamma_k\textbf{D}_k^{\prime}\right)\textbf{w}_k^{\prime}
\end{equation}
where $\textbf{w}_k^{\prime} = [\textbf{w}_{kn_1}^T \: \textbf{w}_{kn_2}^T \: \cdots \: \textbf{w}_{kn_r}^T]^T$ and $\hcap_k^{\prime}, \: \textbf{D}_k^{\prime}$ are defined as in (\ref{P_l2}). Using Cauchy-Schwarz Inequality \cite{CS}, we can show that 
\begin{equation} \label{App7}
\dfrac{(\textbf{w}_k^{\prime})^H\left(\hcap_k^{\prime}(\hcap_k^{\prime})^H-\gamma_k\textbf{D}_k^{\prime}\right)\textbf{w}_k^{\prime}}{(\textbf{w}_k^{\prime})^H\textbf{w}_k^{\prime}} \leq \lambda_{\text{max}}\left(\hcap_k^{\prime}(\hcap_k^{\prime})^H-\gamma_k\textbf{D}_k^{\prime}\right)
\end{equation}
To satisfy (\ref{App5}) we need $\lambda_{\text{max}}\left(\hcap_k^{\prime}(\hcap_k^{\prime})^H-\gamma_k\textbf{D}_k^{\prime}\right)>0$. If this condition holds, then we obtain that
\begin{equation} \label{App8}
\textbf{w}_k^H\textbf{w}_k = (\textbf{w}_k^{\prime})^H\textbf{w}_k^{\prime} \geq \dfrac{\gamma_k\sigma_k^2}{\lambda_{\text{max}}\left(\hcap_k^{\prime}(\hcap_k^{\prime})^H-\gamma_k\textbf{D}_k^{\prime}\right)}.
\end{equation}
Given the index set $S_k$ for user $k$, the power spent by CP for $k$-th user fronthaul data is equal to $\epsilon_1\log_2(1+\gamma_k)|S_k|$. The power spent by RRHs for $k$-th user is equal to $\epsilon_2\textbf{w}_k^H\textbf{w}_k $ which is lower bounded as shown in (\ref{App8}). Therefore, the minimum value of total power spent for $k$-th user $\mathcal{P}_k$ can be found as
\begin{equation} \label{App9}
\mathcal{P}_k = \min_{S_k}\epsilon_1 \log_2(1+\gamma_k)|S_k|+\epsilon_2\dfrac{\gamma_k\sigma_k^2}{\lambda_{\text{max}}\left(\hcap_k^{\prime}(\hcap_k^{\prime})^H-\gamma_k\textbf{D}_k^{\prime}\right)}
\end{equation} 
We find the value of $\mathcal{P}_k$ by minimizing the right-hand side of (\ref{App9}) over all $S_k \subseteq \{1, 2, \ldots, N\}$.


 



\balance

\begin{IEEEbiography}[{\includegraphics[width=1in,height=1.25in,clip,keepaspectratio]{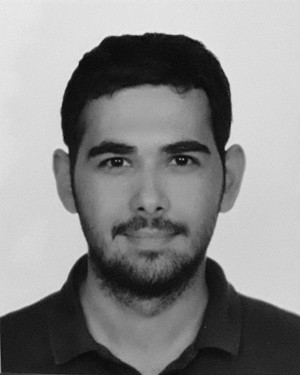}}]{\textbf{Fehm\.{I} Emre Kadan}} received the B.S. and M.S. degrees in electrical and electronics engineering from Middle East Technical University, Ankara, Turkey, in 2013 and 2015, respectively, where he is currently pursuing the Ph.D. degree with the same department. His research interests include wireless communications, signal processing, optimization and beamforming applications.
\end{IEEEbiography}

\begin{IEEEbiography}[{\includegraphics[width=1in,height=1.25in,clip,keepaspectratio]{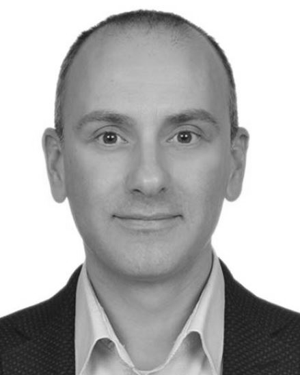}}]{\textbf{Al\.{I} Özgür Yilmaz}} received the B.S. degree in electrical engineering and the M.S. and Ph.D. degrees from the University of Michigan, Ann Arbor, MI, USA, in 1999, 2001, and 2003, respectively. He is currently a Professor with the Department of Electrical and Electronics Engineering, Middle East Technical University, Ankara, Turkey. His research interests include multi-antenna systems, low-complexity transceiver design, effects of transceiver nonidealities on communication systems, coding and information theory in wireless communication systems, 5G communication, resource allocation problems, interference management, and radar signal processing. He has close ties with the industry and contributes to various efforts in developing a local 5G ecosystem.
\end{IEEEbiography}

\EOD

\end{document}